\documentclass[11pt]{article}

\usepackage{amsmath,amssymb,cite}
\usepackage{graphicx,color}

\newcommand{\atopfrac}[2]{\genfrac{}{}{0pt}{}{#1}{#2}}
\newcommand{\f}[2]{\frac{#1}{#2}}

\newcommand{\ko}[1]{\left( #1 \right)}
\newcommand{\kko}[1]{\left[ #1 \right]}
\newcommand{\kkko}[1]{\left\{ #1 \right\}}

\newcommand{\ket}[1]{\left| #1 \right\rangle}

\newcommand{\bmt}[1]{{{\mbox{\boldmath$ #1 $}}}}
\newcommand{\komoji}[1]{\mbox{$#1$}}
\newcommand{\ord}[1]{{\mathcal O\mbox{\small $\left(#1\right)$}}}

\def\w{{\bmt{w}}}

\def\AdSS{{AdS${}_5\times {}$S${}^5$}}
\def\AdSs{{AdS${}_4\times {}$S${}^2$}}
\def\Nf{${\mathcal N}=4$}
\def\H{{\mathcal H}}

\def\N{{\mathcal N}}

\def\eq{\equiv}

\def\al{\alpha}

\def\maru{\mbox{\small $\bigcirc$}}
\def\no{\nonumber}
\def\lam{\lambda}

\def\ep{{\epsilon}}
\def\X{{\mathcal X}}
\def\Y{{\mathcal Y}}
\def\bX{{\overline{\mathcal X}}}
\def\bY{{\overline{\mathcal Y}}}

\numberwithin{equation}{section}
\begin{document}
\quad 
\vspace{-1.0cm}

\begin{flushright}
\parbox{3cm}
{UT-06-04 \hfill \\
KEK-TH-1083 \hfill \\
{\tt hep-th/0604100}\hfill
 }
\end{flushright}

\vspace*{0.5cm}

\begin{center}
\Large\bf 
Higher Loop Bethe Ansatz for Open Spin-Chains in AdS/CFT
\end{center}
\vspace*{0.7cm}
\centerline{\large Keisuke Okamura$^{\dagger,a}$ 
and Kentaroh Yoshida$^{\ddagger,b}$}
\vspace*{0.5cm}
\begin{center}
$^{\dagger}$\emph{Department of Physics, Faculty of Science, \\ 
University of Tokyo, Bunkyo-ku, Tokyo 113-0033, Japan.} 
\\
\vspace*{0.5cm}
$^{\ddagger}$\emph{Theory Division,   	
Institute of Particle and Nuclear Studies, \\
High Energy Accelerator Research 
Organization (KEK),\\
Tsukuba, Ibaraki 305-0801, Japan.} 

\vspace*{0.5cm}
${}^a$ {\tt okamura@hep-th.phys.s.u-tokyo.ac.jp} \qquad  
${}^b$ {\tt kyoshida@post.kek.jp}
\end{center}

\vspace*{0.7cm}

\centerline{\bf Abstract} 

\vspace*{0.5cm}

We propose a perturbative asymptotic Bethe ansatz (PABA) for open
spin-chain systems whose Hamiltonians are given by matrices of anomalous
dimension for composite operators, and apply it to two types of
composite operators related to two different brane configurations. One
is an {\AdSs}-brane in the bulk {\AdSS} which gives rise to a defect
conformal field theory (dCFT) in the dual field theory, and the other is
a giant graviton system with an open string excitation.  In both cases,
excitations on open strings attaching to D-branes (a D5-brane for the
dCFT case, and a spherical D3-brane for the giant graviton case) can be
represented by magnon states in the spin-chains with appropriate
boundary conditions, in which informations of the D-branes are
encoded. We concentrate on single-magnon problems, and explain how to
calculate boundary S-matrices via the PABA technique. We also discuss
 the energy spectrum in
the BMN limit.

\vspace*{1.0cm}

\vfill

\thispagestyle{empty}
\setcounter{page}{0}

\newpage 

\section{Introduction\label{sec:intro}}

The AdS/CFT correspondence \cite{Maldacena:1997re} has continued to give
strong impacts on fundamental aspects of string theories and many
applications to study non-perturbative aspects of gauge theories. It is,
up to now, just a conjecture without rigorous proof, but there are
many evidences to support it without obvious failure. Hence it is
surely an important task to seek more evidences in some tractable
laboratories.  

Recently remarkable developments have been achieved in integrable
structures on both sides of the duality. On the gauge theory side,
Minahan and Zarembo \cite{Minahan:2002ve} showed that a matrix of
anomalous dimension of one-loop planar dilatation operator in the SO(6)
sector is represented by a Hamiltonian of an integrable spin chain. From
their pioneering work, much attention has been payed to integrable
structures in the AdS/CFT correspondence. The classical integrability
of a single system of a superstring on {\AdSS} plays an important role,
and it is considered to be encoded into the integrable structure of the
${\rm PSU}(2,2|4)$ super spin-chain in the dual large $N$ gauge theory
\cite{DNW}. Then the integrable structure is very helpful to study
anomalous dimensions at higher order in perturbation theory, 
since one can use a so-called Bethe ansatz technique in field of
integrable systems.

The Bethe ansatz technique has been developed and utilized in studies of
large $N$ gauge theories very efficiently
\cite{Minahan:2002ve,Beisert:2003xu,Beisert:2003jj,Beisert:2003yb,Beisert:2003ea,Arutyunov1,Arutyunov2,Arutyunov3,Kazakov:2004qf,Serban:2004jf,Schafer-Nameki:2005is,Berenstein:2004ys,Ideguchi:2004wm,Beisert:2005fw,Mann:2005ab,Rej:2005qt,Frolov:2006cc}
(For an exhaustive review, see \cite{Beisert:2004ry}). Based on these
successful examples in closed-strings/closed spin-chains cases, much
efforts have been made to push the ideas toward open-strings/open-chains
cases
\cite{DeWolfe:2004zt,Chen:2004mu,Chen:2004yf,Berenstein:2005vf,McLoughlin:2005gj,Erler:2005nr,Okamura:2005cj,Agarwal:2006gc},
and has seen striking matchings at the one-loop level 
in an effective coupling $\lambda'\equiv g_{\rm YM}^2N/L^2$\,.
Here $L$ is a large quantum number representing the total spin of a string on one side, and the R-charge of an operator super Yang-Mills (SYM) theory on the other.
Such open string sectors are realized by inserting some D-branes or
 orientifolds into the bulk {\AdSS}, and there are many literatures
 dealing with the open string issues
 \cite{DeWolfe:2004zt,Chen:2004mu,Chen:2004yf,Berenstein:2005vf,McLoughlin:2005gj,Erler:2005nr,Okamura:2005cj,Agarwal:2006gc,Berenstein:2002zw,Imamura:2002wz,Balasubramanian:2002sa,Takayanagi:2002nv}.
 There a spin-chain has no longer a periodic boundary condition of a
 closed spin-chain case; instead of the periodicity, it has boundaries
 imposed by the D-brane setup. It is then not clear 
whether the integrability still holds or not with the boundaries. 
In some particular cases, the boundaries were shown to be integrable at least at the one-loop level
\cite{DeWolfe:2004zt,Chen:2004yf,Berenstein:2005vf,Erler:2005nr,
Okamura:2005cj}, i.e., the system can be solved by the Bethe ansatz
technique. When we consider higher-loop contributions to the boundaries,
however, the problem gets more complicated and whether the integrable
structures still remains or not is unclear. One of the aims of this paper
is to examine such an integrability issue.

In comparison to the closed spin-chain case, in the open spin-chain
case, boundary S-matrices appear in Bethe equations as new ingredients.
When the magnon is scattered at either of the left/right boundaries, the
Bethe wavefunction potentially gets a phase shift depending on the
quasi-momentum of the magnon in general. Besides that, the wavefunction
gets phase factors in colliding against other magnons in the bulk of the
chain just as in the closed spin-chain case. The integrability assures
that (i) the magnitude of the momentum does not change, either in
interacting with another magnon, or in reflecting off the boundary, and
(ii) the scattering of spin-waves factorizes into products of
two-body scatterings.  Let the number of the magnons in the open
spin-chain be $M$, and the number of the sites in the chain $L$. Then
the Bethe equation for the open spin-chain typically takes the form
\begin{equation}
\label{Bethe Eq}
e^{2ip_{j}L}=B_{1}(p_{j};g)B_{L}(-p_{j};g)
\prod_{k\neq j,~k=1}^{M}\ko{S_{kj}\ko{p_{k},p_{j};g}S_{kj}\ko{p_{k},-p_{j};g}}
\end{equation}
for $k=1,\dots, M$, where $S_{kj}$ is a two-body bulk S-matrix and
$B_{1,L}$ are boundary S-matrices defined at $x=1$ and $L$\,,
respectively.  When we consider higher-loop corrections, both kinds of
matrices may depend on the gauge coupling $g$ (defined in
(\ref{Hamiltonian2})), which we made explicit in (\ref{Bethe Eq}).
Toward higher loop analysis, the Bethe ansatz technique has to be
generalized so that the S-matrices admit a $g$-expansion. Such a
generalized Bethe ansatz for the closed spin-chain case is often
referred to as \textit{perturbative asymptotic Bethe ansatz (PABA)}. The
PABA technique for the closed spin-chain was established by Staudacher \cite{Staudacher:2004tk} to derive the bulk S-matrix for ${\rm
SU}(1|1)$ sector, and has been utilized in analyzing the integrable structures of 
${\rm SU}(1|2)$ and ${\rm SU}(1,1|2)$ sectors \cite{Beisert:2005fw}, plane-wave matrix model
\cite{Fischbacher:2004iu,Klose:2005cv}.  Recently it was also applied to a
giant graviton system with an open string excitation by Agarwal \cite{Agarwal:2006gc}. 
There the PABA was generalized so that it was
applicable to an open spin-chain with boundaries, and the potential
breakdown of integrability at the two-loop order was suggested.

For the purpose of studying higher-loop contributions to open
spin-chains, our aim here is to examine the integrability of two
distinct open spin-chain models derived 
from two types of composite operators related to two different brane
setups; one is a giant graviton system with an open string excitation,
which is the same system as studied in \cite{Agarwal:2006gc}, and the other is a defect conformal field theory
(dCFT) dual to D3-D5 system. In both models, a bulk two-body S-matrix is
given by the same one as in the four-dimensional $\N=4$ SYM,
and the only object to which we should pay a special attention is the
boundary S-matrix. By analyzing the structure of the boundary S-matrix,
which reflects the field theory structure in concern, we would be able
to draw some valuable information on the string theory side in view of
the AdS/CFT dictionary.

In order to determine the boundary S-matrices, all we have to consider
is
a single-magnon problem, since the scattering at boundaries cannot occur
with more than one magnon. For this reason, in this paper, we focus upon
a single-magnon problem, and leave the problems of potential difficulty
in dealing with more than two magnons at higher loops as a future
work.\footnote{\, A study in this direction has been done for the case
of a giant graviton system in \cite{Agarwal:2006gc}, and it would be
interesting to further examine the problem.}  In a single-trace operator
case, inserting only one magnon in the trace leads to a trivial problem,
but in the present case of ``open'' operators, it is not trivial any
more
due to the presence of boundaries, which indeed makes the problem
essentially a three-body problem, as noticed in \cite{Agarwal:2006gc}.

This paper is organized as follows: In Sec.\,\ref{sec:PABA}, we
propose a PABA for an open spin-chain whose ``bulk'' structure is
determined from the structure of perturbation theory of $\N=4$ SYM.
Then in Sec.\,\ref{sec:GG} and Sec.\,\ref{sec:dCFT}, we will apply it to
two particular models: an open spin-chain on a giant graviton, and an
open spin-chain in dCFT.
Sec.\,\ref{sec:conclusion-discussions} is devoted to a
summary and outlooks.

\section{Perturbative Asymptotic Bethe Ansatz for Open Spin-Chains
\label{sec:PABA}}

In this section, we propose a PABA method for open spin-chains whose
``bulk'' part of the Hamiltonian consists of two complex holomorphic
fields in the $\N=4$ SYM theory, i.e., the fields belonging to an SU(2)
sector. 

\subsection{Open Spin-Chain Hamiltonian from `Open' Diagrams in SYM Theory}

The dilatation operator for single-trace operators in an SU(2)
sector of $\N=4$ SYM can be represented by a Hamiltonian of an
integrable closed spin-chain, and is given by, up to the third order
\cite{Beisert:2003tq,Beisert:2003ys},
\begin{equation}\label{Hamiltonian2}
H_{\rm bulk}=\sum_{r=0}^{\infty}g^{r}\H_{r}\,,\qquad g\eq \f{\lam}{16\pi^{2}}
\end{equation}
where $\lambda\eq g_{\rm YM}^2N$ is the 't Hooft coupling.
The first few Hamiltonian densities $\H_{r}$ are given by
\begin{align}
H_{0}&=\sum_{l=1}^{L}1=L\,,\qquad H_{1}=2\sum_{l=1}^{L-1}Q_{l,l+1}\,,
\nonumber \\
H_{2}&=-8\sum_{l=1}^{L-1}Q_{l,l+1}+2\sum_{l=1}^{L-2}Q_{l,l+2}\,,
\nonumber \\
H_{3}&=60\sum_{l=1}^{L-1}Q_{l,l+1}-24\sum_{l=1}^{L-2}Q_{l,l+2}+4
\sum_{l=1}^{L-3}Q_{l,l+3}\cr
&\qquad {}+b_{1}\sum_{l=1}^{L-3}Q_{l,l+2}Q_{l+1,l+3}+b_{2}
\sum_{l=1}^{L-3}Q_{l,l+3}Q_{l+1,l+2}+b_{3}\sum_{l=1}^{L-3}Q_{l,l+1}Q_{l+2,l+3}
\,.\label{H3}
\end{align}
Here we have defined $Q_{l,k}\eq\f{1}{2}\ko{1-\vec \sigma_{l}\cdot \vec \sigma_{k}}$ with Pauli matrices $\{\vec \sigma_{k}\}_{k=1,2,3}$ for convenience.
The three unknown coefficients $b_{1,2,3}$, which are sensitive to magnon interactions, cannot be determined by just demanding the BMN scaling\footnote{\, If we assume the integrability in
addition to the BMN scaling, however, the coefficients can be fixed as
$b_{1}=4$, $b_{2}=-4$ and $b_{3}=0$ \cite{Beisert:2003tq}. This prediction has recently been
indeed confirmed by a direct perturbative calculation \cite{Eden:2004ua}.}. 
As mentioned in the introduction, we concentrate on a single-magnon
problem, so the $QQ$-terms in (\ref{H3}) are irrelevant to our
present analysis. Hence, by omitting them and rearranging other
$Q$-terms with respect to $k$ in $Q_{l,l+k}$\,, 
the effective ``bulk'' Hamiltonian for the open 
spin-chain can be cast into the following form,
\begin{align}\label{bulk-Hamiltonian}
H_{\rm bulk}'=&\sum_{l=1}^{L-1}c_{1}(g)Q_{l,l+1}
+\sum_{l=1}^{L-2}c_{2}(g)Q_{l,l+2}
+\sum_{l=1}^{L-3}c_{3}(g)Q_{l,l+3}
+\dots\,,
\end{align}
where the coupling-dependent coefficients $c_{k}(g)$ are given by
\begin{align}
c_{1}(g)&=1+2g-8g^{2}+60g^{3}+\dots\,,\\
c_{2}(g)&=2g^{2}-24g^{3}+\dots\,,\\
c_{3}(g)&=4g^{3}+\dots\,.
\end{align}

On the other hand, the ``boundary'' Hamiltonian depends on particular
field theory structures which are determined by brane setups
concerned. We will later see concrete two examples \---- 
one is derived from the giant graviton operators dual to an open string attaching to a giant graviton (Sec.\ \ref{sec:GG}) and the other from the defect 
operator 
dual to an open string attaching to an AdS-brane (D5-brane) in the bulk 
\cite{DeWolfe:2001pq} (Sec.\ \ref{sec:dCFT}).  In both cases, we will
consider an SU(2) sector where the bulk part of the open spin-chains
consists of only two of three kinds of holomorphic 
scalars in $\N=4$ SYM  
(In the dCFT case, in addition to the bulk four-dimensional fields, 
three-dimensional fundamental scalars 
are attached to each end of the composite operators). 

\begin{figure}[htb]
\begin{center}
\vspace{.7cm}
\unitlength 0.1in
\begin{picture}(44.45,26.10)(12.00,-53.15)
%
\special{pn 20}%
\special{pa 2600 3400}%
\special{pa 2600 4800}%
\special{fp}%
%
\special{pn 8}%
\special{pa 2800 3400}%
\special{pa 2800 4800}%
\special{fp}%
%
\special{pn 8}%
\special{pa 3200 3400}%
\special{pa 3200 4800}%
\special{fp}%
%
\special{pn 8}%
\special{pa 3600 3400}%
\special{pa 3600 4800}%
\special{fp}%
%
\special{pn 8}%
\special{pa 4000 3400}%
\special{pa 4000 4800}%
\special{fp}%
%
\special{pn 8}%
\special{pa 4400 3400}%
\special{pa 4400 4800}%
\special{fp}%
%
\special{pn 8}%
\special{pa 4800 3400}%
\special{pa 4800 4800}%
\special{fp}%
\put(30.0000,-46.0000){\makebox(0,0){$\cdots$}}%
%
\special{pn 13}%
\special{pa 3400 4100}%
\special{pa 3400 4800}%
\special{da 0.070}%
%
\special{pn 8}%
\special{pa 4200 4100}%
\special{pa 4200 4800}%
\special{fp}%
\put(38.0000,-46.0000){\makebox(0,0){$\cdots$}}%
\put(46.0000,-46.0000){\makebox(0,0){$\cdots$}}%
%
\special{pn 8}%
\special{pa 3400 3400}%
\special{pa 3400 4000}%
\special{fp}%
%
\special{pn 13}%
\special{pa 4200 3400}%
\special{pa 4200 4000}%
\special{da 0.070}%
%
\special{pn 8}%
\special{sh 0}%
\special{ia 3700 4100 1300 300  0.0000000 6.2831853}%
%
\special{pn 8}%
\special{sh 0.300}%
\special{ia 3700 4100 1300 300  0.0000000 6.2831853}%
%
\special{pn 8}%
\special{pa 2600 5200}%
\special{pa 5600 5200}%
\special{fp}%
%
\special{pn 8}%
\special{pa 5400 3400}%
\special{pa 5400 4800}%
\special{fp}%
%
\special{pn 20}%
\special{pa 5600 3400}%
\special{pa 5600 4800}%
\special{fp}%
%
\special{pn 20}%
\special{sh 1}%
\special{ar 2600 5200 10 10 0  6.28318530717959E+0000}%
\special{sh 1}%
\special{ar 2600 5200 10 10 0  6.28318530717959E+0000}%
%
\special{pn 20}%
\special{sh 1}%
\special{ar 2800 5200 10 10 0  6.28318530717959E+0000}%
\special{sh 1}%
\special{ar 2800 5200 10 10 0  6.28318530717959E+0000}%
%
\special{pn 20}%
\special{sh 1}%
\special{ar 3200 5200 10 10 0  6.28318530717959E+0000}%
\special{sh 1}%
\special{ar 3200 5200 10 10 0  6.28318530717959E+0000}%
%
\special{pn 20}%
\special{sh 1}%
\special{ar 3400 5200 10 10 0  6.28318530717959E+0000}%
\special{sh 1}%
\special{ar 3400 5200 10 10 0  6.28318530717959E+0000}%
%
\special{pn 20}%
\special{sh 1}%
\special{ar 3600 5200 10 10 0  6.28318530717959E+0000}%
\special{sh 1}%
\special{ar 3600 5200 10 10 0  6.28318530717959E+0000}%
%
\special{pn 20}%
\special{sh 1}%
\special{ar 4000 5200 10 10 0  6.28318530717959E+0000}%
\special{sh 1}%
\special{ar 4000 5200 10 10 0  6.28318530717959E+0000}%
%
\special{pn 20}%
\special{sh 1}%
\special{ar 4200 5200 10 10 0  6.28318530717959E+0000}%
\special{sh 1}%
\special{ar 4200 5200 10 10 0  6.28318530717959E+0000}%
%
\special{pn 20}%
\special{sh 1}%
\special{ar 4400 5200 10 10 0  6.28318530717959E+0000}%
\special{sh 1}%
\special{ar 4400 5200 10 10 0  6.28318530717959E+0000}%
%
\special{pn 20}%
\special{sh 1}%
\special{ar 4800 5200 10 10 0  6.28318530717959E+0000}%
\special{sh 1}%
\special{ar 4800 5200 10 10 0  6.28318530717959E+0000}%
%
\special{pn 20}%
\special{sh 1}%
\special{ar 5400 5200 10 10 0  6.28318530717959E+0000}%
\special{sh 1}%
\special{ar 5400 5200 10 10 0  6.28318530717959E+0000}%
%
\special{pn 20}%
\special{sh 1}%
\special{ar 5600 5200 10 10 0  6.28318530717959E+0000}%
\special{sh 1}%
\special{ar 5600 5200 10 10 0  6.28318530717959E+0000}%
%
\special{pn 8}%
\special{sh 0}%
\special{ar 5600 5200 45 45  0.0000000 6.2831853}%
%
\special{pn 8}%
\special{sh 0.600}%
\special{ar 5600 5200 45 45  0.0000000 6.2831853}%
%
\special{pn 8}%
\special{sh 0}%
\special{ar 2600 5200 45 45  0.0000000 6.2831853}%
%
\special{pn 8}%
\special{sh 0.600}%
\special{ar 2600 5200 45 45  0.0000000 6.2831853}%
\put(28.0000,-54.0000){\makebox(0,0){$\X$}}%
\put(32.0000,-54.0000){\makebox(0,0){$\X$}}%
\put(34.0000,-54.0000){\makebox(0,0){$\Y$}}%
\put(36.0000,-54.0000){\makebox(0,0){$\X$}}%
\put(48.0000,-54.0000){\makebox(0,0){$\X$}}%
\put(54.0000,-54.0000){\makebox(0,0){$\X$}}%
%
\special{pn 8}%
\special{pa 2600 3000}%
\special{pa 5600 3000}%
\special{fp}%
%
\special{pn 20}%
\special{sh 1}%
\special{ar 2800 3000 10 10 0  6.28318530717959E+0000}%
\special{sh 1}%
\special{ar 2800 3000 10 10 0  6.28318530717959E+0000}%
%
\special{pn 20}%
\special{sh 1}%
\special{ar 3200 3000 10 10 0  6.28318530717959E+0000}%
\special{sh 1}%
\special{ar 3200 3000 10 10 0  6.28318530717959E+0000}%
%
\special{pn 20}%
\special{sh 1}%
\special{ar 3400 3000 10 10 0  6.28318530717959E+0000}%
\special{sh 1}%
\special{ar 3400 3000 10 10 0  6.28318530717959E+0000}%
%
\special{pn 20}%
\special{sh 1}%
\special{ar 3600 3000 10 10 0  6.28318530717959E+0000}%
\special{sh 1}%
\special{ar 3600 3000 10 10 0  6.28318530717959E+0000}%
%
\special{pn 20}%
\special{sh 1}%
\special{ar 4000 3000 10 10 0  6.28318530717959E+0000}%
\special{sh 1}%
\special{ar 4000 3000 10 10 0  6.28318530717959E+0000}%
%
\special{pn 20}%
\special{sh 1}%
\special{ar 4200 3000 10 10 0  6.28318530717959E+0000}%
\special{sh 1}%
\special{ar 4200 3000 10 10 0  6.28318530717959E+0000}%
%
\special{pn 20}%
\special{sh 1}%
\special{ar 4400 3000 10 10 0  6.28318530717959E+0000}%
\special{sh 1}%
\special{ar 4400 3000 10 10 0  6.28318530717959E+0000}%
%
\special{pn 20}%
\special{sh 1}%
\special{ar 4800 3000 10 10 0  6.28318530717959E+0000}%
\special{sh 1}%
\special{ar 4800 3000 10 10 0  6.28318530717959E+0000}%
%
\special{pn 20}%
\special{sh 1}%
\special{ar 5400 3000 10 10 0  6.28318530717959E+0000}%
\special{sh 1}%
\special{ar 5400 3000 10 10 0  6.28318530717959E+0000}%
%
\special{pn 8}%
\special{sh 0}%
\special{ar 5600 3000 45 45  0.0000000 6.2831853}%
%
\special{pn 8}%
\special{sh 0.600}%
\special{ar 5600 3000 45 45  0.0000000 6.2831853}%
%
\special{pn 8}%
\special{sh 0}%
\special{ar 2600 3000 45 45  0.0000000 6.2831853}%
%
\special{pn 8}%
\special{sh 0.600}%
\special{ar 2600 3000 45 45  0.0000000 6.2831853}%
\put(54.0000,-28.0000){\makebox(0,0){$\bX$}}%
\put(48.0000,-28.0000){\makebox(0,0){$\bX$}}%
\put(44.0000,-28.0000){\makebox(0,0){$\bX$}}%
\put(42.0000,-28.0000){\makebox(0,0){$\bY$}}%
\put(40.0000,-28.0000){\makebox(0,0){$\bX$}}%
\put(28.0000,-28.0000){\makebox(0,0){$\bX$}}%
\put(12.0000,-51.1000){\makebox(0,0)[lt]{$\ket{\rm in}:=\ket{x}$}}%
\put(12.0000,-29.1000){\makebox(0,0)[lt]{$\ket{\rm out}:=\ket{x+s}$}}%
\put(37.1000,-41.0000){\makebox(0,0){boundary interaction}}%
\put(30.0000,-54.0000){\makebox(0,0){$\cdots$}}%
\put(34.0000,-28.0000){\makebox(0,0){$\cdots$}}%
\put(42.0000,-54.0000){\makebox(0,0){$\cdots$}}%
\put(51.0000,-54.0000){\makebox(0,0){$\cdots$}}%
\put(46.0000,-28.0000){\makebox(0,0){$\cdots$}}%
\put(51.0000,-27.9000){\makebox(0,0){$\cdots$}}%
\put(51.0000,-46.0000){\makebox(0,0){$\cdots$}}%
\put(30.0000,-36.0000){\makebox(0,0){$\cdots$}}%
\put(38.0000,-36.0000){\makebox(0,0){$\cdots$}}%
\put(46.0000,-36.0000){\makebox(0,0){$\cdots$}}%
\put(51.0000,-36.0000){\makebox(0,0){$\cdots$}}%
\put(42.0000,-32.0000){\makebox(0,0){$x+s$}}%
\put(34.0000,-50.0000){\makebox(0,0){$x$}}%
\end{picture}%
\vspace{.5cm}
\caption{\footnotesize The diagram for $C_{x,x+s}(g)$\,.  The shaded
area represents the boundary interactions, through which the location of
the impurity changes from $l=x$ to $l=x+s$ (the magnon state is represented by a broken line). This contribution leads to
the dilatation operator for ``open'' composite SYM operators with one
impurity $\Y$\,.
Two gray dots on each ends of the chain stand for the boundary fields.}
\label{fig:C}
\end{center}
\end{figure}
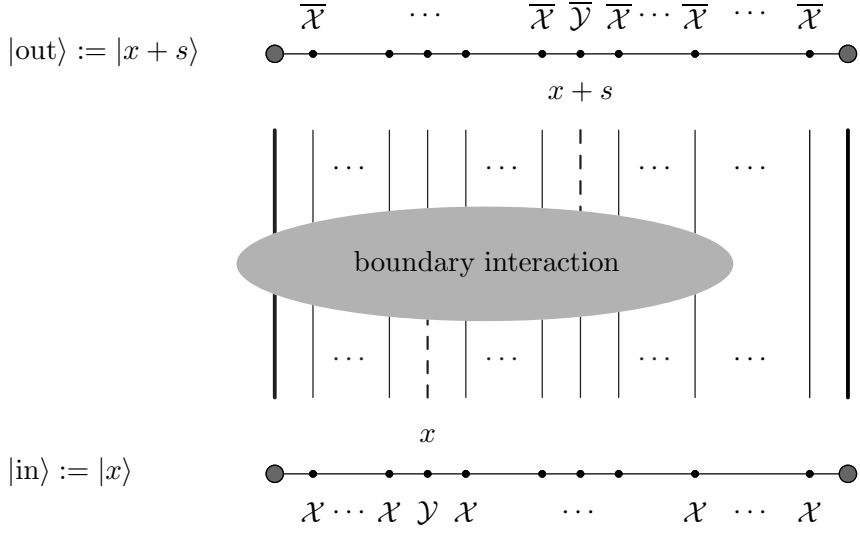

Let the number of the sites in the chain, or the length of the chain,
$L$\,, and label the sites as $x=1, 2, \dots, L$\,. As far as we
consider only one magnon in the chain, the possible structures of Feynman
diagrams can be classified by specifying 
where the magnon starts and
where it ends through the boundary interactions. 
The notation $C_{x,x+s}(g)$ is convenient to indicate the diagrams in
which the magnon starts at site $x$ and end at $x+s$ 
(see Fig.\,\ref{fig:C}).
The change of the location of the magnon, $s$, can either be positive or
negative, but $1\leq x+s\leq L$ must be satisfied for the magnon not to
go out of the spin-chain.  
The boundary Hamiltonian is obtained by summing over all possible diagrams,
\begin{align}\label{boundary-Hamiltonian}
H_{\rm boundary}=\sum C_{x,x+s}(g)\,.
\end{align}
Note that, because of the Hermiticity of the Hamiltonian, if $C_{i,j}$ is contained in (\ref{boundary-Hamiltonian}), so is $C_{j,i}$, and these two give rise to the same contributions.

Now we should ask: what diagrams can contribute to $C_{x,x+s}(g)$ up to some fixed order in $g$?
The following two points should be taken into account.  (i) The higher order
in perturbation we proceed, the wider region around each boundary the
magnon would feel as a scattering area.  (ii) If the magnon in its
in-state locates at $l=x$, only more than $|s|$-loop effect can take it
to $l=x+s$ in its out-state.  
With these in mind, we expect the boundary
Hamiltonian to be given by
\begin{align}
(\ref{boundary-Hamiltonian})&=\sum_{k}^{} g^{k}C_{x,x+s}^{(k)}\cr
&=g\ko{C_{1,1}^{(1)}+C_{L,L}^{(1)}}
+g^{2}\ko{C_{1,1}^{(2)}+C_{1,2}^{(2)}+C_{2,1}^{(2)}+C_{2,2}^{(2)}+C_{L,L}^{(2)}+C_{L,L-1}^{(2)}+C_{L-1,L}^{(2)}+C_{L-1,L-1}^{(2)}}+{}\cr
&\quad {}+g^{3}\left(C_{1,1}^{(3)}+C_{1,2}^{(3)}+C_{1,3}^{(3)}+C_{2,1}^{(3)}+C_{2,2}^{(3)}+C_{2,3}^{(3)}+C_{3,1}^{(3)}+C_{3,2}^{(3)}+C_{3,3}^{(3)}+C_{L,L}^{(3)}+C_{L,L-1}^{(3)}+{}\right.\cr
&{}+{}\left.C_{L,L-2}^{(3)}+C_{L-1,L}^{(3)}+C_{L-1,L-1}^{(3)}+C_{L-1,L-2}^{(3)}+C_{L-2,L}^{(3)}+C_{L-2,L-1}^{(3)}+C_{L-2,L-2}^{(3)}\right)+\ord{g^{4}}\,.
\label{expand:C}
\end{align}
As in (\ref{expand:C}), each $C_{x,x+s}(g)$ will be perturbatively
expanded in $g$.  Our strategy here is to investigate the structure of
boundary S-matrices at higher loops without fixing the coefficients
$C_{x,x+s}^{(k)}$\,, so that it can be used to discuss various open
spin-chain models. 
The total Hamiltonian of
the open spin-chain is then given by the sum of the bulk
(\ref{bulk-Hamiltonian}) and the boundary part
(\ref{boundary-Hamiltonian}),
\begin{equation}
\label{Hamiltonian}
H\eq H_{\rm bulk}'+H_{\rm boundary}\,.
\end{equation}
Up to the third order, by acting the bulk Hamiltonian
(\ref{bulk-Hamiltonian}) and the boundary Hamiltonian
(\ref{boundary-Hamiltonian}) to single-magnon sates $\ket{x}$
$\ko{x=1,2,3,\dots}$\,, we obtain 
\begin{align}
H\ket{1}&=c_{1}(g) \ko{\ket{1}-\ket{2}}
+c_{2}(g) \ko{\ket{1}-\ket{3}}
+c_{3}(g) \ko{\ket{1}-\ket{4}}+{}\cr
&\qquad {}+\ko{gC_{1,1}^{(1)}+g^{2}C_{1,1}^{(2)}+g^{3}C_{1,1}^{(3)}}\ket{1}
+\ko{g^{2}C_{1,2}^{(2)}+g^{3}C_{1,2}^{(3)}}\ket{2}
+g^{3}C_{1,3}^{(3)}\ket{3}
+\ord{g^{4}}\,,\\
H\ket{2}&=c_{1}(g) \ko{-\ket{1}+2\ket{2}-\ket{3}}
+c_{2}(g) \ko{\ket{2}-\ket{4}}
+c_{3}(g) \ko{\ket{2}-\ket{5}}+{}\cr
&\qquad {}
+\ko{g^{2}C_{2,1}^{(2)}+g^{3}C_{2,1}^{(3)}}\ket{1}
+\ko{g^{2}C_{2,2}^{(2)}+g^{3}C_{2,2}^{(3)}}\ket{2}
+g^{3}C_{2,3}^{(3)}\ket{3}
+\ord{g^{4}}\,,\\
H\ket{3}&=c_{1}(g)\ko{-\ket{2}+2\ket{3}-\ket{4}}
+c_{2}(g) \ko{-\ket{1}+2\ket{3}-\ket{5}}
+c_{3}(g) \ko{\ket{3}-\ket{6}}+{}\cr
&\qquad {}+g^{3}C_{3,1}^{(3)}\ket{1}
+g^{3}C_{3,2}^{(3)}\ket{2}
+g^{3}C_{3,3}^{(3)}\ket{3}
+\ord{g^{4}}\,,\\
H\ket{4}&=c_{1}(g)\ko{-\ket{3}+2\ket{4}-\ket{5}}
+c_{2}(g) \ko{-\ket{2}+2\ket{4}-\ket{6}}
+c_{3}(g) \ko{-\ket{1}+2\ket{4}-\ket{7}}+{}\cr
&\qquad {}+\ord{g^{4}}\,,
\end{align}
and so on. In the above expressions, the coefficients $c_{k}(g)$ contain the terms up to
$\ord{g^{3}}$\,; explicitly,
\begin{equation}
c_{1}(g)=g c_{1}^{(1)}+g^{2} c_{1}^{(2)}+g^{3}
c_{1}^{(3)}\,, \qquad 
c_{2}(g)=g^{2} c_{2}^{(2)}+g^{3} c_{2}^{(3)}\,, \qquad 
c_{3}(g)=g^{3} c_{3}^{(3)}\,. \qquad 
\end{equation}
The dispersion relation can be determined only through bulk information,
and is given by, up to $n$-loop order,
\begin{align}
E(p,g)&=c_{1}(g)\ko{2-e^{ip}-e^{-ip}}
+c_{2}(g)\ko{2-e^{2ip}-e^{-2ip}}+\cdots\cr
&\qquad \cdots+c_{n}(g)\ko{2-e^{nip}-e^{-nip}}+\ord{g^{n+1}}\cr
&\eq\sum_{k=1}^{n}\ep_{k}(p)g^{k}+\ord{g^{n+1}}\,.
\end{align} 
Note that we have not yet expanded $p$ in $g$ for the time being, though it may depend on $g$\,. With our Hamiltonian (\ref{Hamiltonian}),
the first few energy coefficients turn out to be,
\begin{equation}
\ep_{1}(p)= 8\sin^{2}\ko{\f{p}{2}}\,,\quad 
\ep_{2}(p)= -32\sin^{4}\ko{\f{p}{2}}\,,~\dots~\,. 
\end{equation}
Our next task is to solve 
an eigenvalue problem, and it will be 
discussed in the next subsection.

\subsection{Bethe Wavefunction and Boundary S-matrix}

In the SU(2) sector the ``bulk'' part of the open spin-chain
consists of two kinds of complex holomorphic 
scalars. Let us denote them as
${\mathcal X}$ and ${\mathcal Y}$, and the letter is regarded as 
the magnon field.   
For the case of an SU(2) open spin-chain derived from a giant graviton 
operators (Sec.\,\ref{sec:GG}), ${\mathcal Y}$ will be identified with the scalar field that form
 a determinant-like operator (i.e., a giant graviton on the gravity side).\footnote{\, This is not the only way to take an SU(2) sector; there is another possibility, in which the magnon state corresponds to an excitation in a Neumann direction to the giant graviton.}
For the case of an SU(2) open spin-chain derived from a defect operator (Sec.\,\ref{sec:dCFT}), on the other hand, 
${\mathcal Y}$ will be a holomorphic complex scalar field in the ${\rm SO}(3)$ vectormultiplet.

For a magnon at site $x$, the eigenvalue equation becomes
\begin{equation}
H|\Psi\rangle=E|\Psi\rangle
\end{equation}
with the Hamiltonian (\ref{Hamiltonian}) and the trial eigenfunction
\begin{align}
&\atopfrac{\hspace{0.3cm}\atopfrac{x}{\downarrow}}
{|\Psi (p)\rangle=\sum\limits_{1\leq x\leq L} \psi(x)\, 
\ket{{\mathcal X}\dots {\mathcal X} {\mathcal Y}{\mathcal X}\dots {\mathcal X}}\eq \sum\limits_{1\leq x\leq L} \psi(x)\, 
\ket{x}\,,}
\end{align}
where we have defined the state with one impurity, or a magnon, ${\mathcal Y}$ at $x$ as $\ket{x}$.
We propose the following ansatz for the Bethe wavefunction on the open spin-chain:
\begin{align}\label{PABA}
\psi (x)&=
\Big(1+f\ko{\left|x-1\right|, p; g}+f\ko{\left|L-x\right|, p; g}\Big)A(p;g)\, e^{ipx}-{}\cr
&\qquad {}-\Big(1+\widetilde f\ko{\left|x-1\right|, -p; g}+\widetilde f\ko{\left|L-x\right|, -p; g}\Big)\widetilde A(-p;g)\, e^{-ipx}\,.
\end{align}
Here the correction function $f\ko{\left| d\right| ,p;g}$ is given by
\begin{equation}\label{f}
f\ko{\left| d\right|,p;g}=\sum_{k=1}^{\infty}f_{k}\ko{\left| d\right|,p}g^{\left| d\right|+k}
+\ord{g^{n+1}}\,,
\end{equation}
and the similar for ${\widetilde f}\ko{\left| d\right| ,p;g}$.
In (\ref{f}), the correction factors $f\ko{\left| x-1\right|,p;g}$ and $f\ko{\left| L-x\right|,p;g}$ 
tend to vanish 
as the magnon moves far away from the boundaries.
This means that the magnon would not feel the long-range interaction with boundaries if the magnon-wave locates sufficiently far away from both ends, in which case the wavefunction reduces to
\begin{equation}\label{psi-0}
\psi (x)\approx A(p;g)\, e^{ipx}-\widetilde A(-p;g)\, e^{-ipx}\eq \psi_{0}(x)\,.
\end{equation}
This is the same ansatz as used in the one-loop analyses of \cite{DeWolfe:2004zt,Berenstein:2005vf,Erler:2005nr,Okamura:2005cj}.
The reader should remind that the quasi-momentum $p$ may also get a $g$-correction, and may be expanded as $p(g)=p^{(0)}+gp^{(1)}+g^{2}p^{(2)}+\dots$\,. 
Another remark is that our PABA (\ref{PABA}) looks slightly different from 
the one used in \cite{Agarwal:2006gc}, but up to the two-loop level, it results n the same boundary S-matrices as in \cite{Agarwal:2006gc}, 
if applied to the giant graviton system.

Now let us explain the origin of the ansatz (\ref{PABA}). 
It is motivated by the PABA for closed spin-chain cases.
As introduced in Sec.\,\ref{sec:intro}, the PABA technique has been used in various situations, and in the work \cite{Agarwal:2006gc}, it was applied to an 
open spin-chain which describes an open string on a giant graviton. 
Recall the PABA for the closed spin-chain case, first introduced by Staudacher \cite{Staudacher:2004tk}.
The ansatz on a wavefunction for a two-magnon state takes the form
\begin{align}\label{PABA-closed}
\psi (x_{1},x_{2})&=\big(1+F\ko{|x_{2}-x_{1}|,p_{1},p_{2};g}\big) A\ko{p_{1},p_{2};g}\, e^{i\ko{p_{1}x_{1}+p_{2}x_{2}}}+{}\cr
&\qquad {}+\big(1+{\widetilde F}\ko{|x_{2}-x_{1}|,p_{2},p_{1};g}\big) {\widetilde A}\ko{p_{2},p_{1};g}\, e^{i\ko{p_{2}x_{1}+p_{1}x_{2}}}\,,
\end{align}
where the correction factor is given by
\begin{equation}\label{f-closed}
F\ko{|d|,p_{1},p_{2};g}=\sum_{k=1}^{\infty} g^{|d|+k} F_{k}(|d|,p_{1},p_{2})\,.
\end{equation}
The bulk two-body S-matrix is defined by $S\ko{p_{2},p_{1};g}\eq {\widetilde A}\ko{p_{2},p_{1};g}/A\ko{p_{1},p_{2};g}$, and admits a $g$-expansion as follows:
\begin{equation}
S\ko{p_{1},p_{2};g}=\sum_{k=0}^{\infty}g^{k}S^{(k)}\ko{p_{1},p_{2}}\,.
\end{equation}
It can be seen that our ansatz (\ref{PABA}) for a single-magnon wavefunction is 
constructed so 
that, at each endpoint, it describes a scattering between a magnon in question at $x_{1}=x$ with $p_{1}=p$ and another ``magnon'' at $x_{2}=1$ (or $x_{2}=L$) with $p_{2}=0$, i.e., the endpoint.
Thus we can regard 
(\ref{PABA}) as a natural extension of the PABA for a closed spin-chain case, (\ref{PABA-closed}) with (\ref{f-closed}), to the open spin-chain case, with the two boundaries playing roles of another two rest ``magnons''.

When we consider the $n$-loop analysis, 
by acting the Hamiltonian (\ref{Hamiltonian}) 
to the Bethe wavefunction with $1+n\leq x\leq L-n$\,,  
we have 
the following eigenvalue equation,
\begin{align}\label{bulk-cond}
\hspace{-0.8cm}\psi (x)\kko{\ep_{1}(p)g^{1}+\ep_{2}(p)g^{2}+\cdots +\ep_{n}(p)g^{n}}&=
c_{1}(g)\ko{2\psi (x)-\psi (x-1)-\psi (x+1)}+{}\cr
&\quad {}+c_{2}(g)\ko{2\psi (x)-\psi (x-2)-\psi (x+2)}+\cdots\cr
&\quad \cdots +c_{n}(g)\ko{2\psi (x)-\psi (x-n)-\psi (x+n)}\,.
\end{align}
On the other hand, when we act the Hamiltonian to a state with the magnon at $x=1,\dots, n$ or $x=L-n+1,\dots, L$, some of the terms in the right-hand side of (\ref{bulk-cond}) do not appear. 
Furthermore the boundary terms in (\ref{Hamiltonian}) give rise to extra terms in the eigenvalue equation.
The compatibility conditions that the bulk condition (\ref{bulk-cond}) persists even at these special points $x=1,\dots, n$ or $x=L-n+1,\dots, L$ lead to a set of Bethe equations.

In what follows, we will mainly concentrate on the scattering problem at the left endpoint $(x=1)$; the analysis on the right end $(x=L)$ can be done in 
the same manner.
The boundary S-matrix is defined as
\begin{equation}\label{def-B}
B(p;g)\eq \f{A(-p;g)}{\widetilde A(p;g)}\,,
\end{equation}
and will be expanded as
\begin{equation}\label{expand-B}
B(p;g)=\sum_{k=0}^{n}B^{(k)}\ko{p}g^{k}+\ord{g^{n+1}}\,.
\end{equation}
As mentioned just before, special care has to be payed 
to 
$x=1$, $2$, $\dots$ $n$ cases, where the boundary terms will affect the eigenvalue equations.
The compatibility conditions lead to a set of constraints among coefficients $\{ f_{k} \}$ and $ \{ {\widetilde f}_{k} \}$ $(k=1,\dots, n-1)$.
Using those equations, 
$B$ can be expressed in terms of the quasi-momentum $p$ in the ansatz (\ref{PABA}) and unfixed coefficients $C^{(k)}_{x,x+s}$ in the boundary Hamiltonian (\ref{boundary-Hamiltonian}).

Let us see how the PABA (\ref{PABA}) works in the concrete case of $n=2$, i.e., the two-loop analysis. 
Compatibility conditions are calculated as, for $x=1$,
\begin{align}\label{cond-x=1}
2\psi_{0} (0)-\big(2-C_{1,1}^{(1)}\big)\psi_{0} (1)
&=g\Big[\big(-2+\ep_{1}(p)-C_{1,1}^{(1)}\big)\,e^{ip}f_{1}(p)B(-p;g)+{}\cr
&\quad \qquad {}+\big(2-\ep_{1}(p)+C_{1,1}^{(1)}\big)\,e^{-ip}{\widetilde f}_{1}(-p) -{}\cr
&\quad \qquad {}-2\psi_{0} (-1)+8\psi_{0} (0)-\big(6+C_{1,1}^{(2)}\big)\psi_{0} (1)-C_{1,2}^{(2)}\psi_{0} (2)\Big]\,,
\end{align}
and for $x=2$,
\begin{equation}\label{cond-x=2}
2B(-p;g)\,e^{ip}f_{1}(p)-2e^{-ip}{\widetilde f}_{1}(-p)-2\psi_{0} (0)-C_{2,1}^{(2)}\psi_{0}(1)-\big(-2+C_{2,2}^{(2)}\big)\psi_{0} (2)=0\,.
\end{equation}
We have similar conditions for $x=L-1$ and $x=L$, and the compatibility conditions at $x=3, 4, \dots, L-2$ become trivial. 
We can solve (\ref{cond-x=2}) for ${\widetilde f}_{1}$, then plugging it into (\ref{cond-x=1}) along with (\ref{psi-0}) (the definition of $\psi_{0}$), we are left with one $p$-dependent condition written in term of $B$ and $C$'s.
Solving it for $B$ and expanding 
it as (\ref{expand-B}), we can determine the coefficients of the series $B^{(0)}$ and $B^{(1)}$ perturbatively. 
Hence 
the resulting $B^{(k)}$ are functions of a quasi-momentum $p$ and the 
undetermined coefficients $C$'s in the Hamiltonian.
We should note that 
$p(g)$ will further allow 
a series expansion in $g$, as we will see later.

Going to the three-loop order ($n=3$), we face a curious situation.
The compatibility condition at $x=3$ together with (\ref{cond-x=2}) leads to
\begin{equation}\label{cond-x=3}
\ko{-2C_{2,1}^{(2)}+C_{3,1}^{(3)}}\psi_{0}(1)+\ko{4+C_{3,2}^{(3)}-2C_{2,2}^{(2)}}\psi_{0}(2)+\ko{-4+C_{3,3}^{(3)}}\psi_{0}(3)=0\,,
\end{equation}
which can be at all satisfied at least for $C_{3,3}^{(3)}=4$; 
otherwise the value of the quasi-momentum would be forced to take 
special values, which would make no sense.
As we will see in the next section,
an open spin-chain related to a giant graviton case is indeed shown to be endowed with $C^{(3)}_{3,3} = 4$, so there is no problem about
 the validity of our Bethe ansatz at least up to three-loop level.
As for the dCFT case, on the other hand,
we have no existing perturbative calculation beyond the one-loop order,
 so we have no positive evidence for the consistency.
But assuming the integrability of the model, in the dCFT case also, one
 may expect that (2.29) is again satisfied trivially with a nice set of
 coefficients.

In the rest part of the paper, we will apply the PABA introduced in this
section to two concrete examples.

\section{PABA for Open Spin-Chains in Giant Graviton System\label{sec:GG}}

Let us first consider an open spin-chain description of an open string
on a giant graviton. There are numerous number of papers concerning the
system; for some of them, see
\cite{Corley:2001zk,Berenstein:2004kk,deMelloKoch:2004ws,deMelloKoch:2005jg,Balasubramanian:2004nb,Berenstein:2003ah}.
A giant graviton is a rotating spherical D3-brane which is wrapping an
S$^{3}$ $\subset$ S$^{5}$ of the {\AdSS} background, and a giant graviton
maximally enhanced in S$^5$ is especially called a maximal giant
graviton.  This is a BPS object described by a determinant-like operator
made up of a single complex scalar field in $\N=4$ SYM, say $Z$.  The
determinant-like operator is called a giant graviton operator.  For the
maximal giant graviton case we can write it as $\ep^{j_{1}\dots
j_{N}}_{i_{1}\dots i_{N}} Z^{i_{1}}_{j_{1}}\dots Z^{i_{N}}_{j_{N}}$,
where the symbol $\ep^{j_{1}\dots j_{N}}_{i_{1}\dots i_{N}}$ is
understood as a product of antisymmetric tensors $\ep^{j_{1}\dots
j_{N}}\ep_{i_{1}\dots i_{N}}$\,.  Attaching an ``open string'' with
length-$(L+2)$ to the maximal giant gives an operator of the form
\begin{equation}\label{open string on GG}
\ep^{j_{1}\dots j_{N}}_{i_{1}\dots i_{N}}Z^{i_{1}}_{j_{1}}\dots Z^{i_{N-1}}_{j_{N-1}}
\ko{M_{0}M_{1}\dots M_{L}M_{L+1}}^{i_{N}}_{j_{N}}\,,
\end{equation}
with the matrix product $M_{0}\dots M_{L+1}$ describing the open string.
We consider the SU(2) sector, where each $M_{l}$ can be either $Z$ or $Y$ 
that are two of the three complex holomorphic scalars in $\N=4$ SYM.

One of the important constraints in our analysis is that we have a boundary condition such that neither $M_{0}$ nor $M_{L+1}$ can be $Z$, since otherwise the operator (\ref{open string on GG}) factorizes into a maximal giant and a closed string, which drops in the large-$N$ analysis 
\cite{Berenstein:2005vf,Berenstein:2003ah}.
We can define the Bethe reference state, or the ground state, as the highest weight state under the unbroken SO(4).
At this time 
we can take it as
\begin{equation}
\ep^{j_{1}\dots j_{N}}_{i_{1}\dots i_{N}} 
Z^{i_{1}}_{j_{1}}\dots Z^{i_{N-1}}_{j_{N-1}}
\ko{Y^{L+2}}^{i_{N}}_{j_{N}}\,.
\end{equation}
Let us now consider a single-magnon excitation on the chain.
We concentrate on the operators of the type
\begin{equation}
\atopfrac{\hspace{0.85cm}\atopfrac{x}{\downarrow}}
{
\ket{x}\eq 
Y\big[\underbrace{Y\dots Y Z Y\dots Y\vphantom{a_{b}}}_{L}\big]Y\,.
}
\end{equation}
with one $Y$ at $l=x$ in (\ref{open string on GG}) replaced 
with $Z$.
As discussed before, 
the boundary condition indicates $x\neq 0$, $L+1$, which in turn implies vanishing of the Bethe wavefunction of $Z$ in the vicinities of $x=0$, $L+1$.

To find the boundary coefficients $C_{x,x+s}^{(k)}$, let us split the first term in (\ref{Hamiltonian}) into three pieces:
\begin{equation}
\sum_{l=0}^{L}c_{1}(g)Q_{l,l+1}=c_{1}(g)\kko{Q_{0,1}+\sum_{l=1}^{L-1}Q_{l,l+1}+Q_{L,L+1}}\,.
\end{equation}
As done above, 
we have to treat $Q_{0,1}$ and $Q_{L,L+1}$ with special care, since they are the only possible terms in the sum 
relevant to the boundary conditions. 
We should be careful in acting these two pieces to the length-($L+2$) word $\w=M_{0}M_{1}\dots M_{L}M_{L+1}$ with $M_{0}=Y$ and $M_{L+1}=Y$
because of the boundary conditions. 
For example, in the case that $Q_{0,1}$ acts on $\w$\,, 
\begin{align}
Q_{0,1}\w=\ko{I-P}_{0,1}\kko{Y M_{1}M_{2}\dots M_{L+1}}=\ko{Y M_{1}-M_{1} Y}M_{2}\dots M_{L+1}\,,
\end{align}
but the state with $M_{0}=Z$ is suppressed in the large $N$ limit, 
so the expression above reduces to $Y Z M_{2}\dots M_{L+1}$ if $M_{1}=Z$.
Or else, if $M_{1}=Y$, we see that $Q_{0,1}\w$ vanishes.
The similar is true for the action of $Q_{L,L+1}$ to $\w$.
Thus we can write 
\begin{equation}
\kko{\sum_{l=0}^{L}c_{1}(g)Q_{l,l+1}}\w
=c_{1}(g)\sum_{l=1}^{L-1}Q_{l,l+1}\w
+c_{1}(g)\ko{q_{1}^{Z}+q_{L}^{Z}}\w\,,
\end{equation}
where we have introduced a projection operator $q_{k}^{X}$ which gives one when the field on the $k$-th site is $X$, and otherwise zero.
In the same way, we can express the action of the second term in (\ref{Hamiltonian}) on $\w$ as 
\begin{equation}
\kko{\sum_{l=0}^{L-1}c_{2}(g)Q_{l,l+2}}\w
=c_{2}(g)\sum_{l=1}^{L-2}Q_{l,l+2}\w
+c_{2}(g)\ko{q_{2}^{Z}+q_{L-1}^{Z}}\w\,,
\end{equation}
In these ways, we can write down the action of the Hamiltonian to general state $\w$ composed of two holomorphic scalars, whose first and the last sites 
cannot be a magnon $Z$, in terms of operators $Q_{l,l+k}$ and
$q_{k}^{Z}$\,. 
This is the case even for 
a general $M$-magnon problem; we have only to add as many boundary terms of the form $q_{k}^{Z}$ as the number of magnons, only two of which can be relevant to the boundary interaction at this two-loop order.

We should keep in mind that this feature we have seen so far is limited to the analysis up to two-loop level.
Three-loop interaction introduces $QQ$-term as in (\ref{H3}), which can 
change the array of magnons between in- and out-states.
For example, in the $M=2$ case, when we begin with the in-state with magnons at $l=1$ and $l=2$, which we denote as $\ket{1,2}$, the operation of $Q_{0,1}Q_{2,3}$ mixes $\ket{1,2}$ and $\ket{1,3}$ in the out-state.
Symbolically, $QQ\ket{x_{1},x_{2}}$ is 
in general not 
proportional to the original word $\ket{x_{1},x_{2}}$ 
due to the boundary interactions.
In such cases, the boundary Hamiltonian cannot be given mealy by projection operators.

Nevertheless, even beyond two-loop analysis, as long as we consider a single-magnon problem, the $QQ$-term never affect the array in the word, and so we can write, for example, the action of the third term in (\ref{Hamiltonian}) as 
\begin{equation}
\kko{\sum_{l=0}^{L-2}c_{2}(g)Q_{l,l+3}}\w
=c_{3}(g)\sum_{l=1}^{L-3}Q_{l,l+3}\w
+c_{3}(g)\ko{q_{3}^{Z}+q_{L-2}^{Z}}\w\,.
\end{equation}

Let us return to the $M=1$ case and concentrate on the two-loop analysis.
In this case, as mentioned before, acting the boundary Hamiltonian to a single-magnon state $\ket{x}$ gives only a state proportional to $\ket{x}$, i.e., we have 
\begin{alignat}{7}
&\text{coeff.\ of\, $g^{1}$:}&\quad
&C_{1,1}^{(1)}=2\,;
&{}&{}&{}&{}\label{GG-C111}\\
&\text{coeff.\ of\, $g^{2}$:}&\quad
&C_{1,1}^{(2)}=-8\,,\quad 
&C_{1,2}^{(2)}=0\,,\quad
&C_{2,1}^{(2)}=0\,,\quad
&C_{2,2}^{(2)}=2\,;\quad
&{}\label{GG-C2}\\
&\text{coeff.\ of\, $g^{3}$:}&\quad
&C_{1,1}^{(3)}=60\,,\quad 
&C_{1,2}^{(3)}=0\,,\quad
&C_{1,3}^{(3)}=0\,,\quad
&C_{2,1}^{(3)}=0\,,\quad
&C_{2,2}^{(3)}=-24\,,\cr
&{}&\quad 
&C_{2,3}^{(3)}=0\,,\quad
&C_{3,1}^{(3)}=0\,,\quad
&C_{3,2}^{(3)}=0\,,\quad
&C_{3,3}^{(3)}=4\,.\quad
&{}\label{GG-C3}
\end{alignat}
Note that the compatibility condition (\ref{cond-x=3}) is satisfied within the coefficients (\ref{GG-C2}) and (\ref{GG-C3}).
Following the procedure explained in Sec.\,\ref{sec:intro}, the boundary
S-matrix $B(p;g)$ defined at $x=0$ can be obtained perturbatively. 

The first series coefficient turns out to be
\begin{align}\label{comp-0}
B^{(0)}(p)&=\f{2-e^{-ip}\big(2-C_{1,1}^{(1)}\big)}{2-e^{ip}\big(2-C_{1,1}^{(1)}\big)}\,,
\end{align}
and thus we can see (\ref{GG-C111}) indeed ensures $B^{(0)}=1$.
The compatibility condition at
$x=L+1$ together with (\ref{comp-0}) gives,
\begin{equation}\label{mom-cond}
1=e^{2ip\ko{L+1}}\cdot \f{2-e^{-ip}\big(2-C_{1,1}^{(1)}\big)}{2-e^{ip}\big(2-C_{1,1}^{(1)}\big)}\cdot \f{2-e^{-ip}\big(2-C_{L,L}^{(1)}\big)}{2-e^{ip}\big(2-C_{L,L}^{(1)}\big)}\,,
\end{equation}
which leads to, at the one-loop order, the following momentum
quantization condition
\begin{equation}
p=\f{n\pi}{L+1}+\ord{g^{1}}\,,
\end{equation}
where we have used the fact $C_{1,1}^{(1)}=2$ and $C_{L,L}^{(1)}=2$\,.
Here $n$ is an integer which labels a momentum index.
Thus the Bethe
wavefunction is given by
\begin{equation}
|\Psi(p)\rangle=a_{n}\sum\limits_{1\leq x\leq L} \sin\kko{\f{\pi n}{L+1}\,x}\ket{x}\,,
\end{equation}
with an $n$-dependent normalization factor $a_{n}$\,.  We can see this wavefunction
indeed vanishes at the endpoints $x=0$ and $x=L+1$, and it gives us an
intuitive picture of an open string attaching to a giant graviton, which
is subject to a Dirichlet boundary condition.
These one-loop analyses were first done by Berenstein and Vazquez \cite{Berenstein:2005vf}. 
Does this matching of boundary conditions still hold at higher order in 
perturbation theory? Or, put it differently, 
can we find the dictionary to relate certain directions in
the string target spacetime to certain fields in the SYM theory, even at
the higher-loop level? 
This is our initial motivation to start the PABA analyses
of open spin-chains resulting from the SYM theory.

As shown below, when we consider higher-loop corrections to the boundary
S-matrices, the boundary S-matrices defined at $x=0$, $L+1$ indeed 
cease to be one due to the higher-loop contributions. In view of AdS/CFT
correspondence, it would be then reasonable to regard somewhere near
$x=0$, $L+1$, at which the wavefunction vanishes, 
as a perturbatively corrected 
definition of the endpoints. This redefinition of the endpoints
provides 
a nice picture even at higher-loop level comparable to the behavior
of an open string attaching to a giant graviton, which obeys exactly the
Dirichlet boundary condition in the strong-coupling region. Or we may be 
able to interpret 
the 
perturbative $g$-correction as implications of an 
interaction between an open string and a brane. 
It would be interesting to confirm this from the direct analysis of
the Dirac-Born-Infeld action.

We argue that the locations of the endpoints
should be corrected by a pure phase factor of the form
$e^{i\theta\ko{p;g}}$ perturbatively in $g$, and assume the following
series expansion:
\begin{equation}\label{theta}
\theta\ko{p;g}=\sum_{k=0}^{\infty}\theta_{k}(p)g^{k}\,.
\end{equation}
Then, by defining 
\begin{equation}\label{def:improved-B}
B_{\rm left}(p;g)\eq e^{i\theta(p;g)}B_{0}(p;g)\,,\qquad 
B_{\rm right}(p;g)\eq e^{-i\theta(p;g)}B_{L+1}(p;g)\,,
\end{equation}
the momentum condition can be expressed as
\begin{equation}\label{mom-cond2}
1=e^{2i\kko{p\ko{L+1}-\theta (p;g)}}B_{\rm left}(p;g) B_{\rm right}(-p;g)\,.
\end{equation}
Demanding both $B_{\rm left}(p;g)$ and $B_{\rm right}(p;g)$ to be one,
the series coefficients of the correction factor are determined as
\begin{equation}
\theta_{0}(p)=0\,,\quad 
\theta_{1}(p)=-4\sin p^{(0)}\,,
\end{equation}
where we have expanded the quasi-momentum as
$p(g)=p^{(0)}+gp^{(1)}+g^{2}p^{(2)}+\dots$~.  Solving
$1=\exp\kkko{2i\kko{p\ko{L+1}-\theta (p;g)}}$ perturbatively, we get
\begin{equation}\label{p-GG}
p(g)=\f{n\pi}{L+1}+\f{4g}{L+1}\sin\ko{\f{n\pi}{L+1}}+\ord{g^{2}}\,.
\end{equation}
This is the same result as obtained in \cite{Agarwal:2006gc} earlier.
We see that, in the two-loop analysis, where to identify with an
endpoint has been corrected from $x=0$ of the one-loop analysis to
$x=0-2g\sin p^{(0)}/p^{(0)}$ with $p^{(0)}=n\pi/(L+1)$.  Similarly, the
right endpoint should be identified with $x=L+1+2g\sin p^{(0)}/p^{(0)}$.
The boundary S-matrices $B_{\rm left/right}(p)$ are thus defined at
these perturbatively corrected 
endpoints to give one.

Finally we can calculate the corrected energy of the 
single-magnon state, or the anomalous dimension of the SYM operator
with one impurity, as
\begin{align}\label{finite-size-GG}
E\komoji{\ko{n;\lambda,L}}&=\f{n^{2}\lambda}{8L^{2}}\kko{1-\f{2}{L}+\f{36-n^{2}\pi^{2}}{12L^{2}}}-\f{n^{4}\lambda^{2}}{128L^{4}}-{}\cr
&\qquad {}-\f{n^{2}\lambda^{2}}{16\pi^{2}L^{3}}\ko{1-\f{3}{L}-\f{18-n^{2}\pi^{2}}{3L^{2}}\dots}
+\ord{\lambda^{3}}\,,
\end{align}
where we have used $\lambda=16\pi^{2}g$. The first line of
(\ref{finite-size-GG}) results from $\ep_{1}+\ep_{2}\big|_{p=p^{(0)}}$,
and the second from the $\ord{g}$-correction term in (\ref{p-GG}). This
is somewhat an unexpected result, 
since the latter contribution diverges in the BMN limit, i.e., $L\to
\infty$ with $\lambda/L^{2}$ kept fixed. 

One may suspect that the apparent breakdown of the BMN scaling results from a wrong counting of the
$N$-dependence of the normalization factor for the states $\w$ when studying the region $\sqrt{N} \sim L$\,. As discussed before, actually we neglected all the states whose first or last site is occupied by the
field $Z$, which is the field forming the giant graviton.  
According to the work \cite{Agarwal:2006gc} that studied the $N$-dependence of the normalization carefully by using the Hamiltonian of \cite{Beisert:2003tq}, however, those states with $Z$ at first or last site should only contribute at the subleading order in $1/N$, hence $1/L^{2}$, therefore, they should be irrelevant to the breakdown. 

After this paper, it is argued in \cite{BCV} that the apparent breakdown of the BMN scaling we have seen implies a breakdown of applicability of a Bethe Ansatz, and that the integrability itself as well as the BMN scaling do exist.
It would be interesting to follow the direction of \cite{BCV} to further examine the problem.

\section{PABA for Open Spin-Chains in Defect Conformal Field Theory\label{sec:dCFT}}

Next we turn to a system of an open spin-chain that appears in a 
defect field theory (dCFT).  We are interested in a D3-D5 system, with a
stack of $N$ coincident D3-branes with one D5-brane probe, the setup
first considered by Karch and Randall \cite{Karch:2000gx}. The action of the dCFT dual
to the D3-D5 system was constructed by DeWolfe, Freedman and Ooguri \cite{DeWolfe:2001pq} and the
superconformality in the non-abelian case was 
studied in \cite{Erdmenger:2002ex}. The original $\N=4$ SYM theory contains
six real scalars $X^{i}$ and adjoint Majorana spinors $\lambda^{\al}$
which transform $\bf 6$ and $\bf 4$ of the SO(6) R-symmetry, and there
is also a gauge field $A_{\mu}$.  In the D3-D5 system, 
the D3-branes fill the $0126$-directions, while the probe D5-brane spans
$012345$ (see the table below).
\begin{center}
\begin{tabular}{c|cccccccccc}
{} &$x_{0}$ &$x_{1}$ &$x_{2}$ &$x_{3}$ &$x_{4}$ &$x_{5}$ &$x_{6}$ &$x_{7}$ &$x_{8}$ &$x_{9}$ \\
\hline
D3 &\maru &\maru &\maru &$\times$ &$\times$ &$\times$ &\maru &$\times$ &$\times$ &$\times$  \\
D5  &\maru &\maru &\maru &\maru &\maru &\maru &$\times$ &$\times$ &$\times$ &$\times$ \\
\end{tabular}
\end{center}
There is a three-dimensional defect in $012$-directions, which
introduces a three-dimensional $\N=4$ hypermultiplet in additional to
the bulk four-dimensional hypermultiplet. This defect preserves ${\rm
SO}(3,2)$ subgroup of the four-dimensional conformal group ${\rm
SO}(4,2)$, and breaks the R-symmetry ${\rm SO}(6)$ to ${\rm SO}(3)_{\rm
H}\times {\rm SO}(3)_{\rm V}$.  The bulk fields are decomposed into a
three-dimensional vector multiplet $\{A_k, P_+\lambda^{\alpha}, X_{\rm
V}^A, D_3X_{\rm H}^I\}$ and a three-dimensional hypermultiplet $\{A_3,
P_-\lambda^{\alpha}, X_{\rm H}^I, D_3 X_{\rm V}^A\}$ with $k=0,1,2$,
$A=1,2,3$ and $I=4,5,6$.  The $X_{\rm H}$ and $X_{\rm V}$ are real
scalar fields which are transformed as $(\bf{3}, \bf{1})$ and $(\bf{1},
\bf{3})$ of ${\rm SO}(3)_{\rm H}\times {\rm SO}(3)_{\rm V}$.  The
four-dimensional Majorana spinor is split into three-dimensional
Majorana spinors $P_{\pm}\lambda^{\al}$, where $P_{\pm}$ are projection
operators, and $\lambda^{\al}\mapsto \lambda_{am}$ are arranged to
transform as $(\bf{2}, \bf{2})$.

We are interested in the SU(2) sector, where composite operators are
made up of two holomorphic scalars $Z\eq X_{\rm H}^{1}+iX_{\rm H}^{2}$
and $W\eq X_{\rm V}^{4}+iX_{\rm V}^{5}$, and the latter plays the role
of a magnon. In addition to these fields, the three-dimensional defect
fields $\overline{q}_{1}$ and $q_{2}$ are needed to form a Bethe state
of our concern (see (\ref{dCFT-op})).

First let us examine the structure of the dilatation operator, or the
spin-chain Hamiltonian, for this sector.
As discussed in Sec.\,\ref{sec:intro}, it is given by
\begin{align}
H&=\sum_{l=1}^{L-1}c_{1}(g)Q_{l,l+1}
+\sum_{l=1}^{L-2}c_{2}(g)Q_{l,l+2}
+\sum_{l=1}^{L-3}c_{3}(g)Q_{l,l+3}\\
&\qquad 
\unitlength 0.1in
\begin{picture}(17.89,8.30)(14.11,-13.60)
%
\special{pn 13}%
\special{pa 1907 600}%
\special{pa 1907 1300}%
\special{dt 0.045}%
\special{pa 1907 1300}%
\special{pa 1907 1299}%
\special{dt 0.045}%
%
\special{pn 8}%
\special{sh 0}%
\special{ia 1907 961 120 120  0.0000000 6.2831853}%
%
\special{pn 8}%
\special{sh 0.300}%
\special{ia 1907 961 120 120  0.0000000 6.2831853}%
\put(19.0700,-9.6100){\makebox(0,0){1}}%
%
\special{pn 13}%
\special{pa 2629 600}%
\special{pa 2629 1300}%
\special{dt 0.045}%
\special{pa 2629 1300}%
\special{pa 2629 1299}%
\special{dt 0.045}%
%
\special{pn 8}%
\special{pa 2870 600}%
\special{pa 2870 1300}%
\special{fp}%
%
\special{pn 8}%
\special{sh 0}%
\special{ia 2749 961 241 120  0.0000000 6.2831853}%
%
\special{pn 8}%
\special{sh 0.300}%
\special{ia 2749 961 241 120  0.0000000 6.2831853}%
\put(27.4900,-9.6100){\makebox(0,0){1}}%
\put(22.6800,-9.6100){\makebox(0,0){$+$}}%
\put(15.4600,-9.6100){\makebox(0,0){$+$}}%
\end{picture}%
\label{dCFT-1}\\
&\qquad 
\unitlength 0.1in
\begin{picture}(27.75,7.00)(14.11,-13.00)
%
\special{pn 13}%
\special{pa 1906 600}%
\special{pa 1906 1300}%
\special{dt 0.045}%
\special{pa 1906 1300}%
\special{pa 1906 1299}%
\special{dt 0.045}%
%
\special{pn 8}%
\special{sh 0}%
\special{ia 1906 960 120 120  0.0000000 6.2831853}%
%
\special{pn 8}%
\special{sh 0.300}%
\special{ia 1906 960 120 120  0.0000000 6.2831853}%
\put(19.0600,-9.6000){\makebox(0,0){2}}%
%
\special{pn 13}%
\special{pa 2624 600}%
\special{pa 2624 1300}%
\special{dt 0.045}%
\special{pa 2624 1300}%
\special{pa 2624 1299}%
\special{dt 0.045}%
%
\special{pn 8}%
\special{pa 2863 600}%
\special{pa 2863 1300}%
\special{fp}%
%
\special{pn 8}%
\special{sh 0}%
\special{ia 2746 960 240 120  0.0000000 6.2831853}%
%
\special{pn 8}%
\special{sh 0.300}%
\special{ia 2746 960 240 120  0.0000000 6.2831853}%
\put(27.4600,-9.6000){\makebox(0,0){2}}%
\put(22.6600,-9.6000){\makebox(0,0){$+$}}%
\put(15.4600,-9.6000){\makebox(0,0){$+$}}%
%
\special{pn 13}%
\special{pa 3581 600}%
\special{pa 3581 1300}%
\special{dt 0.045}%
\special{pa 3581 1300}%
\special{pa 3581 1299}%
\special{dt 0.045}%
%
\special{pn 8}%
\special{pa 3821 600}%
\special{pa 3821 1300}%
\special{fp}%
\put(32.2600,-9.6000){\makebox(0,0){$+$}}%
%
\special{pn 8}%
\special{pa 4060 600}%
\special{pa 4060 1300}%
\special{fp}%
%
\special{pn 8}%
\special{sh 0}%
\special{ia 3826 960 360 120  0.0000000 6.2831853}%
%
\special{pn 8}%
\special{sh 0.300}%
\special{ia 3826 960 360 120  0.0000000 6.2831853}%
\put(38.2600,-9.6000){\makebox(0,0){2}}%
\end{picture}%
\label{dCFT-2}\\
&\hspace{0.88cm} {}+\cdots ~\,,\no
\end{align}
where the second and the third lines of the above Hamiltonian represent
the boundary interactions at the one- and the two-loop order,
respectively.  We omit to write the third-order diagram.  The solid and
the dotted lines represent a four-dimensional bulk and a
three-dimensional defect fields, respectively. The one-loop contribution
(\ref{dCFT-1}) 
was calculated in \cite{DeWolfe:2004zt, Okamura:2005cj}, and at this
level, the location of the magnon $W$ does not change, which can be
easily seen from the Feynman diagrams. Thus the resulting expression for
the boundary Hamiltonian can be expressed in terms of projection
operators, and is given by
\begin{equation}
(\ref{dCFT-1})=4g\ko{q_{1}^{W}+q_{L}^{W}}\,.
\end{equation}
i.e., we have $C_{1,1}^{(1)}=4$ in (\ref{expand:C}). 

From the two-loop order, however, the location of $W$ can in general
change between an in-state and an out-state.  The third diagram in
(\ref{dCFT-2}) contains such diagrams, which have a $qqXX$- and a
$4X$-vertex ($X$ is a four-dimensional scalar $Z$ or $W$, and $q$ is a
three-dimensional scalar). Currently we do not know the values of
$C_{1,1}^{(2)}$, $C_{1,2}^{(2)}$, $C_{2,1}^{(2)}$ and $C_{2,2}^{(2)}$,
and to determine them, it is in principle  
necessary to perform direct two-loop order perturbative calculations for
the boundary interaction. But it seems a hard task, so for now, we
decide to leave them as free parameters.

For our purpose to study the boundary S-matrix and the energy spectrum
via the PABA, we shall define a Bethe reference state as
$\overline{q}_{1}Z^{L} q_{2}$, and set the following single-magnon state
in (\ref{PABA}): 
\begin{equation}\label{dCFT-op}
\atopfrac{\hspace{0.8cm}\atopfrac{x}{\downarrow}}
{\ket{x}\eq 
\overline{q}_{1}\underbrace{Z\dots Z W Z\dots Z \vphantom{a_{b}}}_{L} q_{2}\,.}
\end{equation}
In contrast to the case of an open spin-chain on a giant graviton, for
the defect operators (\ref{dCFT-op}), we define the boundary S-matrix at $x=1/2$ and
$x=L+(1/2)$, reflecting the fact that each of defect fields
$\overline{q}_{1}$ and $q_{2}$ has a bare dimension $1/2$\,. Namely, the magnon wave sees an effective length-$(L+1)$ chain.  The
first series coefficient of the boundary S-matrix defined at $x=1/2$ is
calculated as
\begin{align}
B^{(0)}&=\f{2e^{ip}-\big(2-C_{1,1}^{(1)}\big)}{2-e^{ip}\big(2-C_{1,1}^{(1)}\big)}\,,
\end{align}
but since we already know that $C_{1,1}^{(1)}=4$, the right-hand side of
the above equation vanishes, i.e., $B^{(0)}=1$. This is true for 
the right endpoint, too.   
The momentum condition becomes
\begin{equation}\label{mom-cond3}
1=e^{2ipL}\cdot \f{2e^{ip}-\big(2-C_{1,1}^{(1)}\big)}{2-e^{ip}\big(2-C_{1,1}^{(1)}\big)}\cdot
\f{2e^{ip}-\big(2-C_{L,L}^{(1)}\big)}{2-e^{ip}\big(2-C_{L,L}^{(1)}\big)}\,,
\end{equation}
which leads to, at the one-loop order,
\begin{equation}
p=\f{n\pi}{L}+\ord{g^{1}}
\end{equation}
with the mode index $n$\,. 
The Bethe wavefunction is given by
\begin{equation}
|\Psi(p)\rangle=a_{n}\sum\limits_{1\leq x\leq L} \sin\kko{\f{\pi n}{L}\ko{x-\f{1}{2}}}\ket{x}\,,
\end{equation}
which indeed vanishes at the endpoints $x=1/2$, $L+(1/2)$, as expected
from the behavior of the dual open-string attaching to a brane extended
in the $Z$-direction \cite{DeWolfe:2004zt,Okamura:2005cj}.

Let us continue the study to 
higher loops.
The analysis essentially resembles the case of an open
string on a giant graviton in the previous section.  The momentum
condition becomes
\begin{equation}\label{mom-cond4}
1=e^{2i\kko{pL-\theta (p;g)}}B_{\rm left}(p;g) B_{\rm right}(-p;g)\,.
\end{equation}
with the similar definitions of $B_{\rm left/right}(p;g)$ as in (\ref{def:improved-B}).
Demanding, again, $B_{\rm left}(p;g)=1$ and $B_{\rm right}(p;g)=1$, we get 
\begin{align}
\theta_{0}(p)&=0\,,\\
\theta_{1}(p)&=-\f{1}{2}\tan \Big(\f{p^{(0)}}{2}\Big)
\Big[4\ko{1-4\cos (p^{(0)})-\cos (2p^{(0)})}+C_{1,1}^{(2)}+{}\cr
&\hspace{3.0cm} {}+\ko{1+2\cos p^{(0)}}\ko{C_{1,2}^{(2)}+C_{2,1}^{(2)}}+\ko{1+2\cos p^{(0)}}^{2}C_{2,2}^{(2)}\Big]\,.\label{theta1-dCFT}
\end{align}
Solving $1=\exp\kkko{2i\kko{pL-\theta (p;g)}}$ perturbatively, we get
\begin{align}\label{p-dCFT}
p(g)&=\f{n\pi}{L}+\f{g}{2L}\tan\ko{\f{n\pi}{2L}}
\left[4+C_{1,1}^{(2)}+C_{1,2}^{(2)}+C_{2,1}^{(2)}+3C_{2,2}^{(2)}+2\ko{-2+C_{2,2}^{(2)}}\cos\Big(\f{2n\pi}{L}\Big)+{}\vphantom{\cos\ko{\f{2n\pi}{L}}}\right. \cr
&\hspace{5.0cm} \left. {}+2\ko{-8+C_{1,2}^{(2)}+C_{2,1}^{(2)}+2C_{2,2}^{(2)}}\cos\ko{\f{n\pi}{L}}\right]
+\ord{g^{2}}\,.
\end{align}
Recall that we have the condition $C_{1,2}^{(2)}=C_{2,1}^{(2)}$ because
of the Hermiticity of the Hamiltonian. Then the above expression tells
us that, the quasi-momentum does not depend on the gauge coupling if and
only if the coefficients are given by, in addition to $C_{1,1}^{(1)}=4$,
\begin{equation}\label{special-coeff}
C_{1,1}^{(2)}=-14\,,\quad 
C_{1,2}^{(2)}=C_{2,1}^{(2)}=2\quad \mbox{and}\quad 
C_{2,2}^{(2)}=2\,;
\end{equation}
otherwise, they generally receive $g$-corrections as in the giant
graviton case. This set of coefficients leads to vanishing
$\theta_{1}(p)$ in (\ref{theta1-dCFT}), too.  Note also, the last
condition in (\ref{special-coeff}) is compatible with the condition for
our PABA to be valid even at the three-loop order, which we discussed at
the end of Sec.\,\ref{sec:intro}. Of course, up to now, there is no
reason one can believe the perturbative calculation should result in
(\ref{special-coeff}), but these values are very plausible from the viewpoint of the
integrability.
As we will see soon, the condition (\ref{special-coeff}) also turns out a sufficient condition for the system to have a smooth BMN limit.
These points incline us to believe (\ref{special-coeff}) really predicts the result of direct perturbative computation. It is
an interesting task to carry out the two-loop computation directly in
order to check our argument.

Finally, the single-impurity state energy is calculated as
\begin{align}\label{finite-size-dCFT}
E\komoji{\ko{n;\lambda,L}}&=\f{n^{2}\lambda}{8L^{2}}\ko{1-\f{n^{2}\pi^{2}}{12L^{2}}+\f{n^{4}\pi^{4}}{360L^{4}}+\dots} 
-\f{n^{4}\lambda^{2}}{128L^{4}}\ko{1-\f{n^{2}\pi^{2}}{6L^{2}}+\dots}-{}\cr
&\qquad {}-\f{n^{4}\lambda^{2}}{128L^{4}}\ko{1-\f{n^{2}\pi^{2}}{6L^{2}}+\dots}+{}\cr
&\qquad {}+\f{n^{2}\lambda^{2}}{256\pi^{2}L^{3}}\ko{-16+C_{1,1}^{(2)}+3C_{1,2}^{(2)}+3C_{2,1}^{(2)}+9C_{2,2}^{(2)}}-{}\cr
&\qquad {}-\f{n^{4}\lambda^{2}}{3072L^{5}}\ko{-208+C_{1,1}^{(2)}+15C_{1,2}^{(2)}+15C_{2,1}^{(2)}+81C_{2,2}^{(2)}}
+\ord{\lambda^{3}}\,.
\end{align}
In (\ref{finite-size-dCFT}), the first line stands for the one-loop
energy $\ep_{1}(p^{(0)})$, and the following three lines for the
two-loop energy; the first of which resulting from $\ep_{2}(p^{(0)})$,
and the rest depending on the first corrected piece in the momentum,
$p^{(1)}$.  For (\ref{finite-size-dCFT}) to obey the BMN scaling, we see
the third line has to vanish, since it diverges in the BMN limit.
Interestingly, this condition is satisfied within the choice of the
special coefficients (\ref{finite-size-dCFT}). 
Of course, this choice of coefficients then
leads to the vanishing forth line of (\ref{finite-size-dCFT}), making
the $\ord{g^{2}}$-energy totally $p^{(1)}$-independent.

Finite size-corrections for an open spin-chain in the dCFT was also
calculated by McLoughlin and Swanson \cite{McLoughlin:2005gj}, but the
contributions obtained there resulted only from interactions between
more than one magnons, which is clearly different source from ours
(\ref{finite-size-dCFT}). In fact, the authors of
\cite{McLoughlin:2005gj} treated boundary S-matrices trivially, which
would amount to set (\ref{special-coeff}) from the beginning. If one finds,
in the future calculations of the dCFT, that (\ref{special-coeff}) is
not really the case, then one would have to take account of the
finite-size corrections of (\ref{finite-size-dCFT}) in addition to the
contribution calculated in \cite{McLoughlin:2005gj}.

\section{Summary and Outlook\label{sec:conclusion-discussions}}

In this paper, we have proposed a perturbative asymptotic Bethe ansatz for open spin-chain systems
that have been derived from particular field theory structures and symmetries.  
By considering single-magnon problems for the open spin-chains, we have
analyzed the boundary S-matrices using the PABA method. As concrete
examples, we have considered two models; one is related to a giant graviton
operator (\ref{open string on GG}) and the other to a defect operator (\ref{dCFT-op}).

In the giant graviton case, we have evaluated the numerical coefficients
in the boundary Hamiltonian explicitly, which reflect the contributions of higher-loop
diagrams. This result has enabled us to directly calculate higher-loop
corrections to the boundary S-matrix $B$ as well as in the quasi-momentum
$p$, with our PABA ansatz (\ref{PABA}) and (\ref{f}). Then it has been
shown that the higher-loop contribution introduced $g$-dependence in
both $B$ and $p$, just as was shown in \cite{Agarwal:2006gc}. As the
result, the single-magnon energy acquires a divergent piece in the BMN
limit, the physical interpretation of which remains to be clarified. 

On the other hand, in the dCFT case, there has been no known results concerning
 the values of the boundary coefficients.
  Though these should be determined by direct
perturbative computation, we have not tried to carry out the
perturbative calculation in this paper.  
Instead we have performed
the PABA analysis by leaving them arbitrary constants. 
We have found that $B$'s and
$g$, again, depend on $g$ in general. But, in contrast to the giant
graviton case, it is possible to take 
the coefficients 
so that $B$'s and $g$ become independent of
$g$. If the future direct perturbative calculation really reproduces the
special set of coefficients (\ref{special-coeff}), the energy of the
single-magnon state turns out to have a smooth BMN limit,  
but otherwise it diverges in this limit. 
One should, however, bare in mind that the conclusion made above can only be right when the compatibility conditions are fulfilled to make the Bethe ansatz valid.
When the coefficients turned out to take different values from (\ref{special-coeff}) and did not
satisfy the compatibility conditions, the PABA we employed in Sec.\,\ref{sec:GG} is not valid and it cannot be used to study the energy spectrum of the giant graviton system. 

We have also shown in Sec.\,\ref{sec:PABA} that an open spin-chain
system can have a perturbative integrability beyond the two-loop order,
if and only if the conditions $2C_{2,1}^{(2)}-C_{3,1}^{(3)}=0$,
$4+C_{3,2}^{(3)}-2C_{2,2}^{(2)}=0$ and $C_{3,3}^{(3)}=4$ are satisfied
at the same time, see (\ref{cond-x=3}).  The coefficients for the giant
graviton case, (\ref{GG-C2}) and (\ref{GG-C3}), indeed respect these
relations. Hence, as long as a single-magnon problem is concerned, there
is no apparent breakdown of integrability at least up to the three-loop
level, as in the closed spin-chain case. (Here the terminology
``integrable'' means that the system can be solved via the Bethe
ansatz).  However, see \cite{Agarwal:2006gc}, which implied the
non-integrability of the giant-graviton system by studying a two-magnon
state at the two-loop order; the meaning of the ``integrability''
in \cite{Agarwal:2006gc} slightly differs from our present paper, though.

It would be very interesting to study 
quantum corrections on the 
string side, by directly analyzing the
Dirac-Born-Infeld action.  
In \cite{Balasubramanian:2002sa}, a dictionary
between gauge-invariant BMN operators in the $\N=4$ SYM and the dual
open string states in the pp-wave background was proposed by Balasubramanian, Huang, Levi and Naqvi. There the
Bethe reference state corresponds to the light-cone vacuum, and the
excited state created by a creation operator $a_{n}^{\dagger}{}^{Z}$ in
$Z$-direction with the mode number $n$ corresponds to a superposition of
single-impurity $(Z)$ states with appropriate sine phase-factor
insertion. Explicitly, the dictionary reads (dropping the normalization
factors)
\begin{align}
\ket{-1,0}_{\rm l.c.}\quad 
&\longleftrightarrow\quad 
\ep^{j_{1}\dots j_{N}}_{i_{1}\dots i_{N}}Z^{i_{1}}_{j_{1}}\dots Z^{i_{N-1}}_{j_{N-1}}\ko{Y^{L+2}}{}^{i_{N}}_{j_{N}}\,,\label{vac-GG}\\
a_{n}^{\dagger}{}^{Z}\ket{-1,0}_{\rm l.c.}\quad 
&\longleftrightarrow\quad 
\ep^{j_{1}\dots j_{N}}_{i_{1}\dots i_{N}}Z^{i_{1}}_{j_{1}}\dots Z^{i_{N-1}}_{j_{N-1}}\komoji{\ko{\sum_{l=0}^{L+1}\sin\ko{\f{n\pi l}{L+1}}\, Y^{l} Z Y^{L-l+1}}}{}^{i_{N}}_{j_{N}}\,.\label{BMN-GG}
\end{align}
There exists a known dictionary in the dCFT case, too.  In
\cite{Lee:2002cu}, Lee and Park constructed a dictionary between gauge invariant BMN operators in
the dCFT and the dual open string states in the pp-wave background, which reads:
\begin{align}
\ket{0;p^{+}}_{\rm l.c.}\quad 
&\longleftrightarrow\quad 
\overline{q}_{1}Z^{L}q_{2}\,,\label{vac-dCFT}\\
a_{n}^{\dagger}{}^{W}\ket{0;p^{+}}_{\rm l.c.}\quad 
&\longleftrightarrow\quad 
\komoji{\sum_{l=0}^{L}\,\sin\ko{\f{n\pi l}{L}}
\, \overline{q}_{1}Z^{l}WZ^{L-l}q_{2}}\,.\label{BMN-dCFT}
\end{align}
Our PABA analysis has given $g$-corrections to the sine factors in the
right-hand sides of (\ref{BMN-GG}) and (\ref{BMN-dCFT}), that is,
$p^{(0)}=n\pi/(L+1)$ in (\ref{BMN-GG}) and $p^{(0)}=n\pi/L$ in
(\ref{BMN-dCFT}) are corrected to
$p^{(0)}+g p^{(1)}+\ord{g^{2}}$ as in (\ref{p-GG}) and (\ref{p-dCFT}),
respectively. Then it would be interesting to study how the left-hand
sides of the dictionary will receive corrections in the 't Hooft
coupling $\lambda$ and the inverse tension $\alpha'$\,.

We hope the PABA technique for open spin-chains introduced in this paper
could be applied to other examples as well, so that it would enable us to
capture a deeper nature of open spin-chains, and also of open strings 
on a brane, in light of the AdS/CFT correspondence. 

\subsection*{Acknowledgments}

We would like to thank Y.~Imamura, Y.~Nakayama, Y.~Tachikawa and Y.~Takayama for their valuable comments and discussions.
Especially, we offer special thanks to Y.~Imamura for reading the manuscript and giving us illuminating comments.
We are also grateful to Y.~Takayama for pointing out an error in the original version of this paper, concerning the validity of our PABA for the giant graviton system.
The work of K.\,Y.\ is supported in part by JSPS Research Fellowships for Young Scientists.






\begin{thebibliography}{99}

\bibitem{Maldacena:1997re}
  J.~M.~Maldacena,
``The large $N$ limit of superconformal field theories and supergravity,''
  Adv.\ Theor.\ Math.\ Phys.\  {\bf 2}, 231 (1998)
  [Int.\ J.\ Theor.\ Phys.\  {\bf 38}, 1113 (1999)]
  [arXiv:hep-th/9711200]. \quad 
 E.~Witten,
``Anti-de Sitter space and holography,''
 Adv.\ Theor.\ Math.\ Phys.\  {\bf 2}, 253 (1998)
 [arXiv:hep-th/9802150]. \quad 
S.~S.~Gubser, I.~R.~Klebanov and A.~M.~Polyakov,
``Gauge theory correlators from non-critical string theory,''
  Phys.\ Lett.\ B {\bf 428}, 105 (1998) [arXiv:hep-th/9802109].\quad 
  D.~Berenstein, J.~M.~Maldacena and H.~Nastase,
``Strings in flat space and pp waves from {\Nf} super Yang Mills,''
  JHEP {\bf 0204}, 013 (2002)
  [arXiv:hep-th/0202021].\quad 
  S.~S.~Gubser, I.~R.~Klebanov and A.~M.~Polyakov,
``A semi-classical limit of the gauge/string correspondence,''
  Nucl.\ Phys.\ B {\bf 636}, 99 (2002)
  [arXiv:hep-th/0204051].
%
\bibitem{Minahan:2002ve}
  J.~A.~Minahan and K.~Zarembo,
``The Bethe-ansatz for {\Nf} super Yang-Mills,''
  JHEP {\bf 0303}, 013 (2003)
  [arXiv:hep-th/0212208].
%
\bibitem{DNW}
L.~Dolan, C.~R.~Nappi and E.~Witten, 
``A relation between approaches to integrability in superconformal
	Yang-Mills theory,'' JHEP {\bf 0310} (2003) 017 [arXiv:hep-th/0308089].
\bibitem{Beisert:2003xu}
  N.~Beisert, J.~A.~Minahan, M.~Staudacher and K.~Zarembo,
``Stringing spins and spinning strings,''
  JHEP {\bf 0309}, 010 (2003)
  [arXiv:hep-th/0306139].

\bibitem{Beisert:2003jj}
  N.~Beisert,
``The complete one-loop dilatation operator of {\Nf} super Yang-Mills
theory,''
  Nucl.\ Phys.\ B {\bf 676}, 3 (2004)
  [arXiv:hep-th/0307015].
%
\bibitem{Beisert:2003yb}
  N.~Beisert and M.~Staudacher,
``The {\Nf} SYM integrable super spin chain,''
  Nucl.\ Phys.\ B {\bf 670}, 439 (2003)
  [arXiv:hep-th/0307042].
%
\bibitem{Beisert:2003ea}
  N.~Beisert, S.~Frolov, M.~Staudacher and A.~A.~Tseytlin,
``Precision spectroscopy of AdS/CFT,''
  JHEP {\bf 0310}, 037 (2003)
  [arXiv:hep-th/0308117].
%
%
\bibitem{Arutyunov1}
 G.~Arutyunov and M.~Staudacher,
  ``Matching higher conserved charges for strings and spins,''
  JHEP {\bf 0403} (2004) 004
  [arXiv:hep-th/0310182].

\bibitem{Arutyunov2}
  G.~Arutyunov, S.~Frolov, J.~Russo and A.~A.~Tseytlin,
  ``Spinning strings in AdS(5) x S**5 and integrable systems,''
  Nucl.\ Phys.\ B {\bf 671} (2003) 3
  [arXiv:hep-th/0307191].

\bibitem{Arutyunov3}
 G.~Arutyunov, S.~Frolov and M.~Staudacher,
  ``Bethe ansatz for quantum strings,''
  JHEP {\bf 0410} (2004) 016
  [arXiv:hep-th/0406256].

\bibitem{Kazakov:2004qf}
  V.~A.~Kazakov, A.~Marshakov, J.~A.~Minahan and K.~Zarembo,
``Classical / quantum integrability in AdS/CFT,''
  JHEP {\bf 0405}, 024 (2004)
  [arXiv:hep-th/0402207].\quad
  V.~A.~Kazakov and K.~Zarembo,
``Classical / quantum integrability in non-compact sector of AdS/CFT,''
  JHEP {\bf 0410}, 060 (2004)
  [arXiv:hep-th/0410105].\quad 
  N.~Beisert, V.~A.~Kazakov and K.~Sakai,
``Algebraic curve for the $SO(6)$ sector of AdS/CFT,''
  [arXiv:hep-th/0410253].\quad
  N.~Beisert, V.~A.~Kazakov, K.~Sakai and K.~Zarembo,
``The algebraic curve of classical superstrings on {\AdSS},''
  [arXiv:hep-th/0502226].\quad
  S.~Schafer-Nameki,
``The algebraic curve of 1-loop planar {\Nf} SYM,''
  Nucl.\ Phys.\ B {\bf 714}, 3 (2005)
  [arXiv:hep-th/0412254].

\bibitem{Serban:2004jf}
  D.~Serban and M.~Staudacher,
``Planar {\Nf} gauge theory and the Inozemtsev long range spin chain,''
  JHEP {\bf 0406} (2004) 001
  [arXiv:hep-th/0401057].
\bibitem{Schafer-Nameki:2005is}
  S.~Schafer-Nameki and M.~Zamaklar,
``Stringy sums and corrections to the quantum string Bethe ansatz,''
  JHEP {\bf 0510}, 044 (2005)
  [arXiv:hep-th/0509096].\quad 
  S.~Schafer-Nameki, M.~Zamaklar and K.~Zarembo,
``Quantum corrections to spinning strings in AdS(5) x S**5 and Bethe ansatz:
A comparative study,''
  JHEP {\bf 0509}, 051 (2005)
  [arXiv:hep-th/0507189].






\bibitem{Berenstein:2004ys}
  D.~Berenstein and S.~A.~Cherkis,
   ``Deformations of N = 4 SYM and integrable spin chain models,''
  %
  Nucl.\ Phys.\ B {\bf 702}, 49 (2004)
  [arXiv:hep-th/0405215].

\bibitem{Ideguchi:2004wm}
  K.~Ideguchi,
   ``Semiclassical strings on AdS(5) x S**5/Z(M) and operators in orbifold
   field theories,''
  %
  JHEP {\bf 0409}, 008 (2004)
  [arXiv:hep-th/0408014].
  \quad 
  N.~Beisert and R.~Roiban,
   ``The Bethe ansatz for Z(S) orbifolds of N = 4 super Yang-Mills theory,''
  %
  JHEP {\bf 0511}, 037 (2005)
  [arXiv:hep-th/0510209].
\bibitem{Beisert:2005fw}
  N.~Beisert and M.~Staudacher,
   ``Long-range PSU(2,2$|$4) Bethe ansaetze for gauge theory and strings,''
  %
  Nucl.\ Phys.\ B {\bf 727}, 1 (2005)
  [arXiv:hep-th/0504190].

\bibitem{Mann:2005ab}
  N.~Mann and J.~Polchinski,
   ``Bethe ansatz for a quantum supercoset sigma model,''
  %
  Phys.\ Rev.\ D {\bf 72}, 086002 (2005)
  [arXiv:hep-th/0508232].
\bibitem{Rej:2005qt}
  A.~Rej, D.~Serban and M.~Staudacher,
``Planar N = 4 gauge theory and the Hubbard model,''
  JHEP {\bf 0603}, 018 (2006)
  [arXiv:hep-th/0512077].
\bibitem{Frolov:2006cc}
  S.~Frolov, J.~Plefka and M.~Zamaklar,
``The AdS(5) x S**5 superstring in light-cone gauge and its Bethe
equations,''
  arXiv:hep-th/0603008.
\bibitem{Beisert:2004ry}
  N.~Beisert,
``The dilatation operator of {\Nf} super Yang-Mills theory and
integrability,''
  Phys.\ Rept.\  {\bf 405}, 1 (2005)
  [arXiv:hep-th/0407277].
\bibitem{DeWolfe:2004zt}
  O.~DeWolfe and N.~Mann,
   ``Integrable open spin chains in defect conformal field theory,''
  %
  JHEP {\bf 0404}, 035 (2004)
  [arXiv:hep-th/0401041].
\bibitem{Chen:2004mu}
  B.~Chen, X.~J.~Wang and Y.~S.~Wu,
   ``Integrable open spin chain in super Yang-Mills and the plane-wave / SYM
   duality,''
  %
  JHEP {\bf 0402}, 029 (2004)
  [arXiv:hep-th/0401016].\quad 
  B.~J.~Stefanski,
   ``Open spinning strings,''
  %
  JHEP {\bf 0403}, 057 (2004)
  [arXiv:hep-th/0312091].\quad 
  Y.~Susaki, Y.~Takayama and K.~Yoshida,
   ``Open semiclassical strings and long defect operators in AdS/dCFT
   correspondence,''
%
  Phys.\ Rev.\ D {\bf 71}, 126006 (2005)
  [arXiv:hep-th/0410139].\quad 
  Y.~Susaki, Y.~Takayama and K.~Yoshida,
   ``Integrability and higher loops in AdS/dCFT correspondence,''
  %
  Phys.\ Lett.\ B {\bf 624}, 115 (2005)
  [arXiv:hep-th/0504209].\quad 
  D.~Berenstein, D.~H.~Correa and S.~E.~Vazquez,
   ``Quantizing open spin chains with variable length: An example from giant
   gravitons,''
  %
  Phys.\ Rev.\ Lett.\  {\bf 95}, 191601 (2005)
  [arXiv:hep-th/0502172].

\bibitem{Chen:2004yf}
  B.~Chen, X.~J.~Wang and Y.~S.~Wu,
   ``Open spin chain and open spinning string,''
  %
  Phys.\ Lett.\ B {\bf 591}, 170 (2004)
  [arXiv:hep-th/0403004].
  \bibitem{Berenstein:2005vf}
  D.~Berenstein and S.~E.~Vazquez,
``Integrable open spin chains from giant gravitons,''
  JHEP {\bf 0506}, 059 (2005)
  [arXiv:hep-th/0501078].


\bibitem{McLoughlin:2005gj}
  T.~McLoughlin and I.~J.~Swanson,
``Open string integrability and AdS/CFT,''
  Nucl.\ Phys.\ B {\bf 723}, 132 (2005)
  [arXiv:hep-th/0504203].

\bibitem{Erler:2005nr}
  T.~G.~Erler and N.~Mann,
``Integrable open spin chains and the doubling trick in N = 2 SYM with
fundamental matter,''
  JHEP {\bf 0601}, 131 (2006)
  [arXiv:hep-th/0508064].
\bibitem{Okamura:2005cj}
  K.~Okamura, Y.~Takayama and K.~Yoshida,
``Open spinning strings and AdS/dCFT duality,''
  JHEP {\bf 0601} (2006) 112
  [arXiv:hep-th/0511139].
\bibitem{Agarwal:2006gc}
  A.~Agarwal,
``Open spin chains in super Yang-Mills at higher loops: Some potential
problems with integrability,''
  [arXiv:hep-th/0603067].


\bibitem{Berenstein:2002zw}
  D.~Berenstein, E.~Gava, J.~M.~Maldacena, K.~S.~Narain and H.~Nastase,
   ``Open strings on plane waves and their Yang-Mills duals,''
  %
  arXiv:hep-th/0203249.

\bibitem{Imamura:2002wz}
  Y.~Imamura,
   ``Open string: BMN operator correspondence in the weak coupling regime,''
  %
  Prog.\ Theor.\ Phys.\  {\bf 108}, 1077 (2003)
  [arXiv:hep-th/0208079].
\bibitem{Balasubramanian:2002sa}
  V.~Balasubramanian, M.~X.~Huang, T.~S.~Levi and A.~Naqvi,
``Open strings from N = 4 super Yang-Mills,''
  JHEP {\bf 0208}, 037 (2002)
  [arXiv:hep-th/0204196].
\bibitem{Takayanagi:2002nv}
  H.~Takayanagi and T.~Takayanagi,
   ``Notes on giant gravitons on pp-waves,''
  %
  JHEP {\bf 0212}, 018 (2002)
  [arXiv:hep-th/0209160].
\bibitem{Staudacher:2004tk}
  M.~Staudacher,
   ``The factorized S-matrix of CFT/AdS,''
  %
  JHEP {\bf 0505}, 054 (2005)
  [arXiv:hep-th/0412188].

\bibitem{Fischbacher:2004iu}
  T.~Fischbacher, T.~Klose and J.~Plefka,
   ``Planar plane-wave matrix theory at the four loop order: Integrability
   without BMN scaling,''
  %
  JHEP {\bf 0502}, 039 (2005)
  [arXiv:hep-th/0412331].
\bibitem{Klose:2005cv}
  T.~Klose,
``On the breakdown of perturbative integrability in large N matrix models,''
  JHEP {\bf 0510}, 083 (2005)
  [arXiv:hep-th/0507217].
%
\bibitem{Beisert:2003tq}
  N.~Beisert, C.~Kristjansen and M.~Staudacher,
   ``The dilatation operator of N = 4 super Yang-Mills theory,''
  %
  Nucl.\ Phys.\ B {\bf 664}, 131 (2003)
  [arXiv:hep-th/0303060].


\bibitem{Beisert:2003ys}
  N.~Beisert,
``The su(2$|$3) dynamic spin chain,''
  Nucl.\ Phys.\ B {\bf 682}, 487 (2004)
  [arXiv:hep-th/0310252].
%
\bibitem{Eden:2004ua}
  B.~Eden, C.~Jarczak and E.~Sokatchev,
``A three-loop test of the dilatation operator in N = 4 SYM,''
  Nucl.\ Phys.\ B {\bf 712}, 157 (2005)
  [arXiv:hep-th/0409009].
\bibitem{DeWolfe:2001pq}
  O.~DeWolfe, D.~Z.~Freedman and H.~Ooguri,
   ``Holography and defect conformal field theories,''
  %
  Phys.\ Rev.\ D {\bf 66}, 025009 (2002)
  [arXiv:hep-th/0111135].
\bibitem{Corley:2001zk}
  S.~Corley, A.~Jevicki and S.~Ramgoolam,
``Exact correlators of giant gravitons from dual N = 4 SYM theory,''
  Adv.\ Theor.\ Math.\ Phys.\  {\bf 5}, 809 (2002)
  [arXiv:hep-th/0111222].

\bibitem{Berenstein:2004kk}
  D.~Berenstein,
``A toy model for the AdS/CFT correspondence,''
  JHEP {\bf 0407} (2004) 018
  [arXiv:hep-th/0403110].

\bibitem{deMelloKoch:2004ws}
  R.~de Mello Koch and R.~Gwyn,
``Giant graviton correlators from dual SU(N) super Yang-Mills theory,''
  JHEP {\bf 0411}, 081 (2004)
  [arXiv:hep-th/0410236].

\bibitem{deMelloKoch:2005jg}
  R.~de Mello Koch, N.~Ives, J.~Smolic and M.~Smolic,
``Unstable giants,''
  Phys.\ Rev.\ D {\bf 73}, 064007 (2006)
  [arXiv:hep-th/0509007].

\bibitem{Balasubramanian:2004nb}
  V.~Balasubramanian, D.~Berenstein, B.~Feng and M.~x.~Huang,
``D-branes in Yang-Mills theory and emergent gauge symmetry,''
  JHEP {\bf 0503}, 006 (2005)
  [arXiv:hep-th/0411205].

\bibitem{Berenstein:2003ah}
  D.~Berenstein,
``Shape and holography: Studies of dual operators to giant gravitons,''
  Nucl.\ Phys.\ B {\bf 675}, 179 (2003)
  [arXiv:hep-th/0306090].


\bibitem{BCV}
  D.~Berenstein, D.~H.~Correa and S.~E.~Vazquez,
   ``A study of open strings ending on giant gravitons, spin chains and
  integrability,''
  arXiv:hep-th/0604123.

\bibitem{Karch:2000gx}
  A.~Karch and L.~Randall,
   ``Open and closed string interpretation of SUSY CFT's on branes with
   boundaries,''
  %
  JHEP {\bf 0106}, 063 (2001)
  [arXiv:hep-th/0105132].


\bibitem{Erdmenger:2002ex}
  J.~Erdmenger, Z.~Guralnik and I.~Kirsch,
   ``Four-dimensional superconformal theories with interacting boundaries or
   defects,''
  %
  Phys.\ Rev.\ D {\bf 66}, 025020 (2002)
  [arXiv:hep-th/0203020].

%
\bibitem{Lee:2002cu}
  P.~Lee and J.~W.~Park,
   ``Open strings in PP-wave background from defect conformal field theory,''
  %
  Phys.\ Rev.\ D {\bf 67}, 026002 (2003)
  [arXiv:hep-th/0203257].















  










\end{thebibliography}
\end{document}